\documentclass[arps,twocolumn,showpacs]{revtex4}

\usepackage{bm}
\usepackage{graphicx}
\usepackage{subfigure}
\input epsf

\newcommand\lsim{\mathrel{\rlap{\lower4pt\hbox{\hskip1pt$\sim$}}
        \raise1pt\hbox{$<$}}}
\newcommand\gsim{\mathrel{\rlap{\lower4pt\hbox{\hskip1pt$\sim$}}
        \raise1pt\hbox{$>$}}}

\def\thetaB{\mbox{\boldmath$\theta$}}

\def\dthetaB{\delta \mbox{\boldmath$\theta$}}

\begin{document}
\title{Anisotropic Magnification Distortion of the 3D Galaxy Correlation: \\
II. Fourier and Redshift Space}

\author{Lam Hui$^{1,2,3}$, Enrique Gazta\~{n}aga$^{4}$ and Marilena LoVerde$^{1,2}$}

\affiliation{
$^{1}$Institute for Strings, Cosmology and Astroparticle Physics (ISCAP)\\
$^{2}$Department of Physics, Columbia University, New York, NY 10027, U.S.A.\\
$^{3}$Institute of Theoretical Physics, The Chinese University of Hong Kong, Hong Kong\\
$^{4}$Institut de Ci\`encies de l'Espai, IEEC-CSIC, Campus UAB,
F. de Ci\`encies, Torre C5 par-2,  Barcelona 08193, Spain\\
lhui@astro.columbia.edu, gazta@aliga.ieec.uab.es, marilena@phys.columbia.edu
}
\date{\today}

\begin{abstract}
In paper I of this series we discuss how magnification bias distorts the 3D 
correlation function by enhancing the observed correlation
in the line-of-sight (LOS) orientation, especially on large scales.
This lensing anisotropy is distinctive, making it possible to separately measure
the galaxy-galaxy, galaxy-magnification {\it and} magnification-magnification
correlations. Here we extend the discussion to the power spectrum and also
to redshift space.
In real space, pairs oriented close to the LOS direction are not protected against
nonlinearity even if the pair separation is large; this is because
nonlinear fluctuations can enter through gravitational
lensing at a small transverse separation (or i.e. impact parameter).
The situation in Fourier space is different: by focusing on
a small wavenumber $k$, as is usually done, linearity is guaranteed because both
the LOS and transverse wavenumbers must be small.
This is why magnification distortion of the galaxy correlation appears
less severe in Fourier space. Nonetheless, the effect is non-negligible,
especially for the transverse Fourier modes, and should be taken into account
in interpreting precision measurements of the galaxy power spectrum, 
for instance those that focus on the baryon oscillations.
The lensing induced anisotropy of the power spectrum has a shape that is
distinct from the more well known redshift space anisotropies due to peculiar motions
and the Alcock-Paczynski effect. The lensing anisotropy is highly localized
in Fourier space while redshift space distortions are more spread out. This means
that one could separate the magnification bias component in 
real observations, implying that potentially it is possible to perform a gravitational 
lensing measurement  without measuring galaxy shapes.
\end{abstract}



\maketitle

\section{Introduction}
\label{intro}

The effect of magnification bias on the 3D galaxy/quasar
correlation function
was studied in paper I \cite{3Dpaper1}.  
(Galaxy and quasar can be considered synonymous hereafter.)
With the important exception of the classic paper by Matsubara \cite{matsubara00},
previous work on how magnification bias modifies clustering observations
has largely focused on the 2D angular correlation function
\cite{TOG84,J95,BTP95,VFC97,MJV98,MJ98,EGmag03,ScrantonSDSS05,menard,bhuv,JSS03}.
The novelty of the 3D correlation function, as emphasized by
\cite{matsubara00} and \cite{3Dpaper1}, is that magnification bias makes it anisotropic.
In this paper, we extend our previous analysis by studying the anisotropy
in Fourier and redshift space. At first sight, the extension to Fourier
space might seem a trivial exercise.
The calculations are indeed straightforward, but as we will see,
the results are far from obvious: there are important qualitative
differences between the results in Fourier space and real space that 
go beyond the usual wavenumber-position ($k$-$x$) duality.

Let us recall the situation in real space, as depicted in paper I.
The anisotropy of the observed 3D correlation function can be understood
intuitively as follows.
The correlation function is measured by
pair counts of galaxies. A pair of galaxies that are aligned along the 
line-of-sight (LOS) behave differently from a pair oriented transverse to
the LOS. In the former case, the closer galaxy can lens the background one.
The same does not happen in the transverse orientation. The net effect is
an anisotropy in the observed correlation function, induced by
gravitational lensing (or equivalently, magnification bias; we refer to
this effect as magnification distortion).
This reasoning suggests the gravitational lensing
corrections are largest for a pair of galaxies oriented along the
LOS. Indeed, the corrections can be quite significant: consider
for instance a LOS separation of $\sim 100$ Mpc/h;
the intrinsic galaxy correlation is rather weak on such a large scale,
but the lensing induced correction can be quite substantial,
since for the LOS orientation, the relevant lensing impact parameter,
i.e. transverse separation, is zero
(keep in mind also that the lensing effect grows with the LOS separation
while the intrinsic galaxy correlation generally drops with separation).
In other words, for the LOS orientation, scales that otherwise would
be considered {\it linear} can in fact be secretly affected
by nonlinear fluctuations via lensing. 
A large separation $|\delta {\bf x}| = \sqrt{\delta\chi^2 + |\delta {\bf x_\perp}|^2}$ 
does not guarantee linearity
because nonlinear fluctuations can sneak in through lensing with a small transverse
separation $|\delta {\bf x_\perp}|$.

This peculiar mixing of linear intrinsic galaxy fluctuations
with nonlinear lensing fluctuations does not arise in Fourier space.
A small net wavenumber $|{\bf k}|$ is sufficient to guarantee
that both the LOS component $k_\parallel$ and the transverse
component $|{\bf k_\perp}|$ are small. Nonlinear lensing
corrections cannot sneak in as long as one focuses on a small $|{\bf k}|$,
as is usually done. This immediately tells us that the anisotropy
of the observed galaxy correlation must appear milder in Fourier space.
Our primary goal here is to quantify this.

As discussed in paper I, the magnification bias induced anisotropy in the
observed 3D galaxy correlation has two implications.
First, precision measurements of the galaxy correlation must take
into account such magnification distortion.
These include future galaxy surveys that hope to determine the baryon
oscillation scale to high accuracy \cite{eisenstein,2dFa,2dFb,hutsi,tegmark,baotheory,baoexp}.
Second, the distinctive anisotropy pattern makes it possible in principle
to separately measure the galaxy-galaxy, galaxy-magnification {\it and} magnification-magnification
correlations. 
(The last correlation was generally ignored in previous papers that focused
on angular correlation between galaxies at widely separated redshifts
\cite{MJ98,EGmag03,ScrantonSDSS05,menard,bhuv,JSS03}, where
the galaxy-magnification correlation dominates.)
Achieving such a separation requires that one understands other
sources of anisotropy. We therefore extend our
Fourier analysis here to incorporate the anisotropy due to 
both peculiar motions and the Alcock-Paczynski effect.

The rest of the paper is organized as follows.
In \S \ref{distort}, we derive and numerically compute the
magnification distortion 
of the observed galaxy clustering --
\S \ref{correlation} summarizes the results from paper I
 on the correlation function while \S \ref{pk} focuses on the power spectrum.
Redshift space distortion due to peculiar motion is next
incorporated in \S \ref{anisotropy},
and we conclude in \S \ref{discuss}.
In Appendix \ref{app:AP}, we discuss 
the Alcock-Paczynski anisotropy.

Before we start, it is useful to point out several related papers.
Vallinoto et al. \cite{scott} explored the impact of lensing, especially
magnification bias, on the baryon oscillation signal in the real
space correlation function. Their results
are consistent with ours in paper I, though they focus exclusively on pair separations
that are oriented transverse to the LOS, and
their work is therefore more connected to our paper on the angular correlation function
\cite{wLHG}. Wagner et al. \cite{steinmetz} examined the anisotropy
of the 3D correlation that is introduced by light cone effects.
A discussion of the classic paper by Matsubara \cite{matsubara00} can be found
in paper I. Both \cite{matsubara00} and paper I focused on the real/configuration
space correlation function, though peculiar motions and the Alcock-Paczynski 
effect are also treated in \cite{matsubara00}.
A recent paper by Zhang \& Chen \cite{zhangchen} explored the effects of gravitational
lensing in Fourier space in the context of supernova observations.
LoVerde et al. \cite{LHGisw} examined the impact of magnification bias on
integrated Sachs-Wolfe measurements.

\section{Magnification Distortion}
\label{distort}

Given an intrinsic galaxy overdensity $\delta_g$, 
magnification bias introduces a correction  $\delta_\mu$ to the
observed galaxy overdensity $\delta_{\rm obs}$:
\begin{eqnarray}
\label{start}
\delta_{\rm obs} = \delta_g + \delta_\mu
\end{eqnarray}
which is a function of
the galaxy position, specified for instance by the radial comoving distance $\chi$
and the angular position $\thetaB$. 
The magnification bias correction is given by
\cite{TOG84,J95,BTP95,VFC97,MJV98,MJ98}:
\begin{eqnarray}
\label{deltamu}
\delta_\mu = (5s - 2) \kappa
\end{eqnarray}
where $\kappa$ is the lensing convergence:
\begin{eqnarray}
\kappa(\chi,\thetaB) = \int_0^\chi d\chi' {\chi'(\chi-\chi') \over \chi} \nabla^2_\perp \phi(\chi', \thetaB)
\end{eqnarray}
$\phi$ is the gravitational potential, and
$\nabla^2_\perp$ is the 2D Laplacian in the transverse directions. We assume a flat universe -- 
generalization to an open or a closed universe is straightforward. 
The symbol $s$ stands for
\begin{eqnarray}
s = {d {\,\rm log}_{10} N(< m) \over dm}
\end{eqnarray}
where $N(< m)$ is the cumulative number counts for galaxies brighter
than magnitude $m$. This assumes the galaxy sample is defined by
a sharp faint-end cut-off. 
A broader definition of $s$ for a more general galaxy selection is given in Appendix A of
paper I.

We define the galaxy bias $b$ by $\delta_g = b \delta$, where $\delta$ is the mass overdensity.  Eq. (\ref{start}) can then be rewritten as
\begin{eqnarray}
\label{start2}
{\delta_{\rm obs} \over b} = \delta + {5s-2 \over b} \kappa
\end{eqnarray}
The relative importance of the intrinsic clustering and the magnification 
bias correction is therefore controlled by, among other things, the sample
 dependent ratio $(5s-2)/b$. 

The precise values of $s$ and $b$ depend sensitively on details of how the galaxy/quasar
sample is selected, for instance subject to color cuts and so on.
Unless otherwise stated, we adopt throughout this paper the value $(5s-2)/b = 1$ to illustrate
the effect of magnification bias on clustering measurements (see paper I for more details).
An implicit assumption is the linearity of  the galaxy bias, a subject we
will return to in \S \ref{discuss}.

In all illustrative examples below, we employ the following
cosmological parameters:  the Hubble constant $h = 0.7$, matter density 
$\Omega_m = 0.27$, cosmological constant $\Omega_\Lambda = 0.73$, 
baryon density $\Omega_b = 0.046$, 
power spectrum slope $n = 0.95$ and normalization $\sigma_8 = 0.8$.
We employ the transfer function of \cite{EH98}, and 
the prescription of \cite{smith} for the nonlinear power spectrum.
In all equations we use units where the speed of light is unity: $c=1$.

\subsection{The Correlation Function}
\label{correlation}

Here, we summarize the main results of paper I.
Including lensing magnification, the observed two-point correlation function
is given by:
\begin{eqnarray}
\label{obs}
\xi_{\rm obs} (\chi_1, \thetaB_1; \chi_2, \thetaB_2) =
\langle \delta_{\rm obs} (\chi_1, \thetaB_1) \delta_{\rm obs} (\chi_2, \thetaB_2) \rangle \\ \nonumber
= \xi_{gg} (\chi_1, \thetaB_1; \chi_2, \thetaB_2) + \xi_{g\mu} (\chi_1, \thetaB_1; \chi_2, \thetaB_2) \\ \nonumber
+ \xi_{g\mu} (\chi_2, \thetaB_2; \chi_1, \thetaB_1) + \xi_{\mu\mu} (\chi_1, \thetaB_1; \chi_2, \thetaB_2)
\end{eqnarray}
where the magnification bias corrections, 
the galaxy-magnification and magnification-magnification correlations,
are:
\begin{eqnarray}
\label{gmu}
\xi_{g\mu} (\chi_1, \thetaB_1; \chi_2, \thetaB_2) + \xi_{g\mu} (\chi_2, \thetaB_2; \chi_1, \thetaB_1) =
\\ \nonumber 
{3\over 2} H_0^2 \Omega_m (5s - 2) (1+\bar z)
| \chi_2 - \chi_1 | \\ \nonumber
\int {d^2 k_\perp \over (2\pi)^2} P_{gm} (\bar z, k_\perp) 
e^{i {\bf k_\perp} \cdot \bar\chi (\thetaB_1 - \thetaB_2)}
\end{eqnarray}
\begin{eqnarray}
\label{mumu}
\xi_{\mu\mu} (\chi_1, \thetaB_1; \chi_2, \thetaB_2) =
[{3\over 2} H_0^2 \Omega_m (5s - 2)]^2 \\ \nonumber
\int_0^{\bar\chi} d\chi' \left[{(\bar\chi - \chi')\chi' \over \bar\chi}\right]^2 (1+z')^2 \\ \nonumber
\int {d^2 k_\perp \over (2\pi)^2} P_{mm} (z', k_\perp) e^{i {\bf k_\perp} \cdot \chi' (\thetaB_1 - \thetaB_2)}
\end{eqnarray}
Here we have Taylor expanded $\chi_1$ and $\chi_2$ around the mean 
$\bar\chi$ and retained the lowest
order contributions. The
intrinsic (unlensed) galaxy auto-correlation, or galaxy-galaxy correlation,
is 
\begin{eqnarray}
\label{gg}
&& \xi_{gg} (\chi_1, \thetaB_1; \chi_2, \thetaB_2) = \\ \nonumber &&
\xi_{gg} (\sqrt{ (\chi_1-\chi_2)^2 + 
\bar\chi^2 (\thetaB_1 - \thetaB_2)^2}) = \\ \nonumber &&
\int {d^3 k \over (2\pi)^3} P_{gg} (\bar z, k) e^{i {\bf k} \cdot ({\bf x_1} - {\bf x_2})}
\end{eqnarray}
ignoring for now the issue of redshift distortion, 
which will be addressed in \S \ref{anisotropy}. Note that ${\bf x_1}$ and ${\bf x_2}$ refer
to the points corresponding to $\chi_1, \thetaB_1$ and $\chi_2, \thetaB_2$.
Note also $P_{mm}$, $P_{gm}$ and $P_{gg}$ denote respectively the
mass-mass, galaxy-mass and galaxy-galaxy power spectra.

The observed correlation function is a sum of all three correlations above (eq.
[\ref{gmu}], [\ref{mumu}] and [\ref{gg}]). 
(A discussion of their higher order Taylor corrections 
can be found in Appendix B of paper I.).
Viewed in this way, the anisotropy of the lensing induced corrections
is quite striking: $\xi_{g\mu}(1,2) + \xi_{g\mu}(2,1)$ scales linearly with the line-of-sight (LOS) separation
$|\chi_2 - \chi_1|$
(i.e. it increases rather than decreases with the separation!), and 
$\xi_{\mu\mu}$ is independent of the LOS separation.
The intrinsic galaxy auto-correlation
$\xi_{gg}$ is isotropic and generally decreases with separation.

We can summarize the distinctive lensing induced anisotropy in the
observed correlation function as follows:
\begin{eqnarray}
\label{scaling2}
\xi_{\rm obs} (\delta\chi, \delta x_\perp)
= && \xi_{gg} (\sqrt{\delta\chi^2 + \delta x_\perp^2}) \\ \nonumber && 
+ f(\delta x_\perp) \delta\chi
+ g(\delta x_\perp)
\end{eqnarray}
where $\delta\chi$ and $\delta x_\perp$ are the LOS and transverse separations
respectively, $f \delta\chi$ represents the galaxy-magnification correlation
and $g$ represents the magnification-magnification correlation.
Here, $f$ and $g$ are functions of the transverse separation only, and
are determined by the galaxy-mass and mass-mass power spectra.
{\it This distinctive form of the anisotropy allows us in principle
to separately measure $\xi_{gg}$, $f$ and $g$, from which we can
infer the galaxy-galaxy, galaxy-mass and mass-mass power spectra.}
For instance, at any given $\delta x_\perp$, plotting 
$\xi_{\rm obs}$ as a function of the LOS separation $\delta\chi$ would reveal a
linear contribution at sufficiently large $\delta\chi$'s
where $\xi_{gg}$ is very small. Its slope tells us
$f$ and its extrapolation to $\delta\chi=0$ tells us $g$. 
Subtracting $\delta\chi f + g$ from $\xi_{\rm obs}$ then yields $\xi_{gg}$.
This is illustrated in Fig. 2 of paper I, where
we also present order of magnitude estimates for the ratios
$\xi_{\mu\mu}/\xi_{gg}$ and $\xi_{g\mu}/\xi_{gg}$:
\begin{eqnarray}
\label{xiorder}
{\xi_{\mu\mu}\over \xi_{gg}} \sim \left[5s-2 \over b\right]^2 
{(1+ \bar z)^2\over 50}(\bar\chi H_0)^3 {\pi H_0\over k_*} {\Delta^2 (k_*) \over
\Delta^2 (k_{**})}  \\ \nonumber
{2\xi_{g\mu} \over \xi_{gg}} \sim  \left[5s-2 \over b\right] {1+\bar z \over 2}
(\delta\chi H_0) {\pi H_0\over k_*} {\Delta^2 (k_*) \over
\Delta^2 (k_{**})}
\end{eqnarray}
The symbol $\Delta^2 (k)$ denotes the
dimensionless variance at scale $k$ and redshift $\bar z$: $4\pi k^3 P_{mm} (k)/(2\pi)^3$. 
Here, $k_{**} \sim 1/\sqrt{\delta\chi^2 + \delta x_\perp^2}$,
while $k_*$ is equal to either $1/\delta x_\perp$ or $k_m$, whichever
is smaller ($k_m$ is the scale where $k^2 P_{mm} (k)$ peaks; $k_m \gsim 3$ h/Mpc).
These estimates work reasonably well except for separations around or
beyond the zero-crossing scale.

For the LOS orientation, where $\Delta^2 (k_*)$ can be much larger
than $\Delta^2 (k_{**})$, the magnification bias corrections
$\xi_{\mu\mu}$ and $\xi_{g\mu}$ can dominate over the intrinsic clustering
correlation $\xi_{gg}$. The implications for baryon acoustic oscillation
measurements are summarized at the end of \S \ref{pk}.

\subsection{The Power Spectrum}
\label{pk}

For surveys with a simple geometry, the power spectrum is often the
more popular quantity to measure.
Suppose the survey geometry is specified by $W({\bf x}) = W_\parallel 
(x_\parallel) W_\perp({\bf x_\perp})$,
where ${\bf x}$ specifies a location with $x_\parallel$ being the LOS
component and ${\bf x_\perp}$ the transverse component. For instance, a top-hat geometry
in the radial direction is described by
\begin{eqnarray}
\label{tophat}
W_\parallel (x_\parallel) &=& 1/\sqrt{L} \quad {\rm if} \quad -L/2 < x_\parallel < L/2 \\ \nonumber
&=& 0 \quad {\rm otherwise}
\end{eqnarray}
with $L$ being the radial span. A Gaussian geometry in the radial direction would
be described by
\begin{eqnarray}
\label{gauss}
W_\parallel (x_\parallel) &=&  (\pi \sigma^2)^{-1/4} {\rm exp\,} [ -{x_\parallel^2 /(2 \sigma^2)} ]
\end{eqnarray}
Our normalization convention is that $\int d^3 x W^2 = 1$. 
Note that the origin $x_\parallel = 0$ is chosen to be located at the mean redshift
of interest.

The Fourier counterparts of eq. (\ref{gmu}) and (\ref{mumu}), taking into account
the effects of the window, are
\begin{eqnarray}
\label{Pgmu}
&& 2 P_{g\mu} ({\bf k}) = {3\over 2} H_0^2 \Omega_m (5s - 2) (1+\bar z) \\ \nonumber
&& \quad G(k_\parallel) \int {d^2 k_\perp' \over (2\pi)^2} P_{gm} (\bar z, k_\perp') |\tilde W_\perp ({\bf k_\perp} -
{\bf k_\perp'})|^2
\end{eqnarray}
and
\begin{eqnarray}
\label{Pmumu}
P_{\mu\mu} ({\bf k}) = [{3\over 2} H_0^2 \Omega_m (5s - 2)]^2 
|\tilde W_\parallel (k_\parallel)|^2 \\ \nonumber
\int_0^{\bar\chi} d\chi' (\bar\chi - \chi')^2 (1+z')^2 \\ \nonumber
\int {d^2 k_\perp' \over (2\pi)^2} P_{mm} (z', 
k_\perp' \bar\chi/\chi') |\tilde W_\perp ({\bf k_\perp} -
{\bf k_\perp'}) |^2
\end{eqnarray}
where $\tilde W_\parallel$ and $\tilde W_\perp$ are Fourier transforms of the
windows $W_\parallel$ and $W_\perp$, and $G$ is defined as follows:
\begin{eqnarray}
\label{Gdef}
G(k_\parallel) = \int dx_1 dx_2 |x_1 - x_2| 
e^{ik_\parallel (x_1 - x_2)} W_\parallel(x_1) W_\parallel (x_2)
\end{eqnarray}
where $x_1$ and $x_2$ represent the LOS distance.

The galaxy power spectrum is windowed in the usual way:
\begin{eqnarray}
\label{Pgg}
P_{gg} ({\bf k}) = \int {d^3 k' \over (2\pi)^3} P_{gg}^{\rm true} (k') |\tilde W({\bf k - \bf k'})|^2
\end{eqnarray}
where $\tilde W$ is the Fourier transform of the total window $W$, and 
$P_{gg}^{\rm true}$ represents the true/unwindowed galaxy power spectrum.

The observed power spectrum is the sum:
\begin{eqnarray}
\label{Pobs}
P_{\rm obs} ({\bf k}) = P_{gg} ({\bf k}) + 2 P_{g\mu} ({\bf k}) + P_{\mu\mu} ({\bf k})
\end{eqnarray}
where $P_{\rm obs} ({\bf k})$ is defined to be
\begin{eqnarray}
P_{\rm obs} ({\bf k}) = \int d^3 x_1 d^3 x_2 \langle \delta_{\rm obs} ({\bf x_1}) \delta_{\rm obs} ({\bf x_2}) \rangle \\ \nonumber W({\bf x_1}) W({\bf x_2}) e^{i{\bf k} \cdot ({\bf x_1} - {\bf x_2})}
\\ \nonumber
\end{eqnarray}
The survey window enters into these three contributions to the observed power in distinct ways.
For the galaxy-galaxy power spectrum, the window function is convolved as usual under an integral
with the true 3D power. 
For $P_{\mu\mu}$, only the transverse window function is  convolved under an integral
with the power spectrum ($P_{mm}$). The LOS window function
is not convolved under an integral at all: $P_{\mu\mu}$ is directly
proportional to $|\tilde W_\parallel |^2$. For $P_{g\mu}$, it is also directly proportional to
some generalized LOS window function $G(k_\parallel)$. This interesting behavior of
the magnification-magnification and galaxy-magnification power spectra can be traced to the unique anisotropies
of their real space counterparts: referring back to eq. (\ref{scaling2}),
the appearance of $|\tilde W_\parallel |^2$ in $P_{\mu\mu}$ is related to the
fact that $g$ is independent of the LOS separation, and
the appearance of $G$ in $P_{g\mu}$ is related to the linear dependence of
$f \delta\chi$ on the LOS separation.

The precise form of the window functions depends on the exact geometry. For the example of the top-hat 
window (eq. [\ref{tophat}]),
they are
\begin{eqnarray}
\label{tophatFT}
&& |\tilde W_\parallel (k_\parallel) |^2 = {4 \over L k_\parallel^2} [{\,\rm sin\,}(k_\parallel L/2)]^2
\\ \nonumber
&& G(k_\parallel) = {2 \over k_\parallel^2} 
\left[ {2 \over k_\parallel L} {\,\rm sin\,} (k_\parallel L) - {\,\rm cos\,}(k_\parallel L) - 1\right]
\end{eqnarray}
At low $k_\parallel$, $|\tilde W_\parallel |^2 \sim L$ while
$G \sim L^2/3$.

\begin{figure}[tb]
\centerline{\epsfxsize=9cm\epsffile{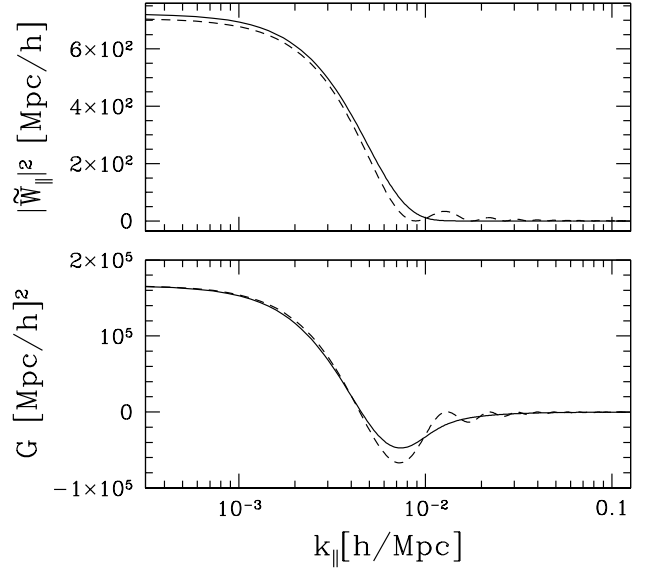}}
\caption{The multiplicative LOS windows $|\tilde W_\parallel|^2$ and
$G_\parallel$ as a function of the LOS wavenumber $k_\parallel$. 
The dashed lines are for a top-hat geometry, with $L = 706.6$ Mpc/h and
the solid lines are for a Gaussian geometry, with $\sigma = 204$ Mpc/h.
}
\label{gwinoutT}
\end{figure}

The corresponding expressions for the example of the Gaussian window (eq. [\ref{gauss}]) are
\begin{eqnarray}
\label{gaussFT}
&& |\tilde W_\parallel (k_\parallel) |^2 = \sqrt{4 \pi \sigma^2} e^{- \sigma^2 k_\parallel^2} \\ \nonumber
&& G(k_\parallel) = 4\sigma^2 - 8 k_\parallel^2 \sigma^4 \int_{0}^1 dy e^{k_\parallel^2 \sigma^2 (y^2 - 1)}
\end{eqnarray}
At low $k_\parallel$, $|\tilde W_\parallel |^2 \sim \sqrt{4\pi} \sigma$, while
$G \sim 4\sigma^2$. Illustrations of $|\tilde W_\parallel|^2$ and $G$ can be found
in Fig. \ref{gwinoutT} for both the top-hat and Gaussian geometries.

\begin{figure*}[tb]
\subfigure[]
{\label{pkcontour.new.try.1}\includegraphics[width=.45\textwidth]{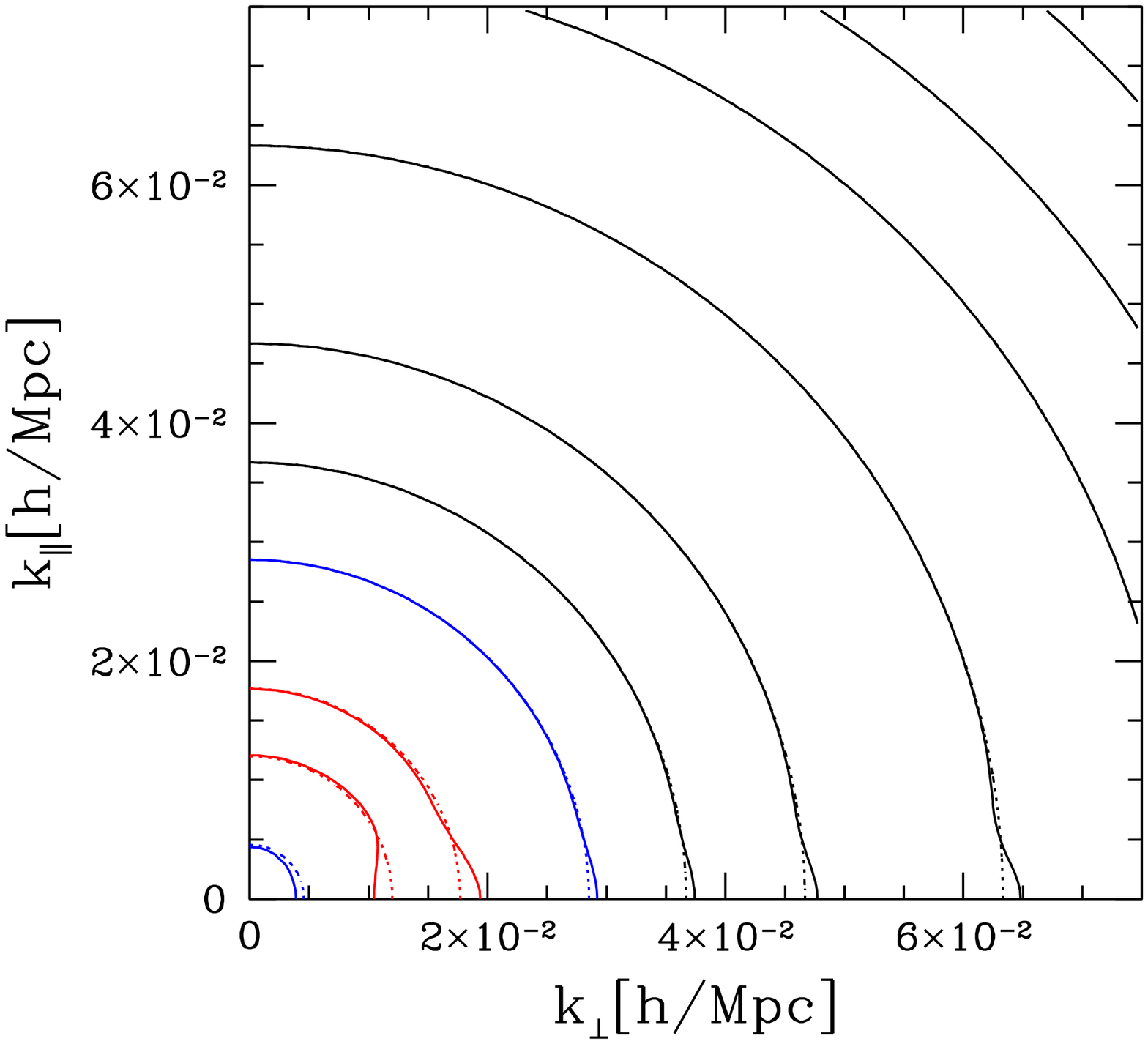}}
\hspace{0.1in}
\subfigure[]
{\label{pkBBKS2.trye2.1}\includegraphics[width=.45\textwidth]{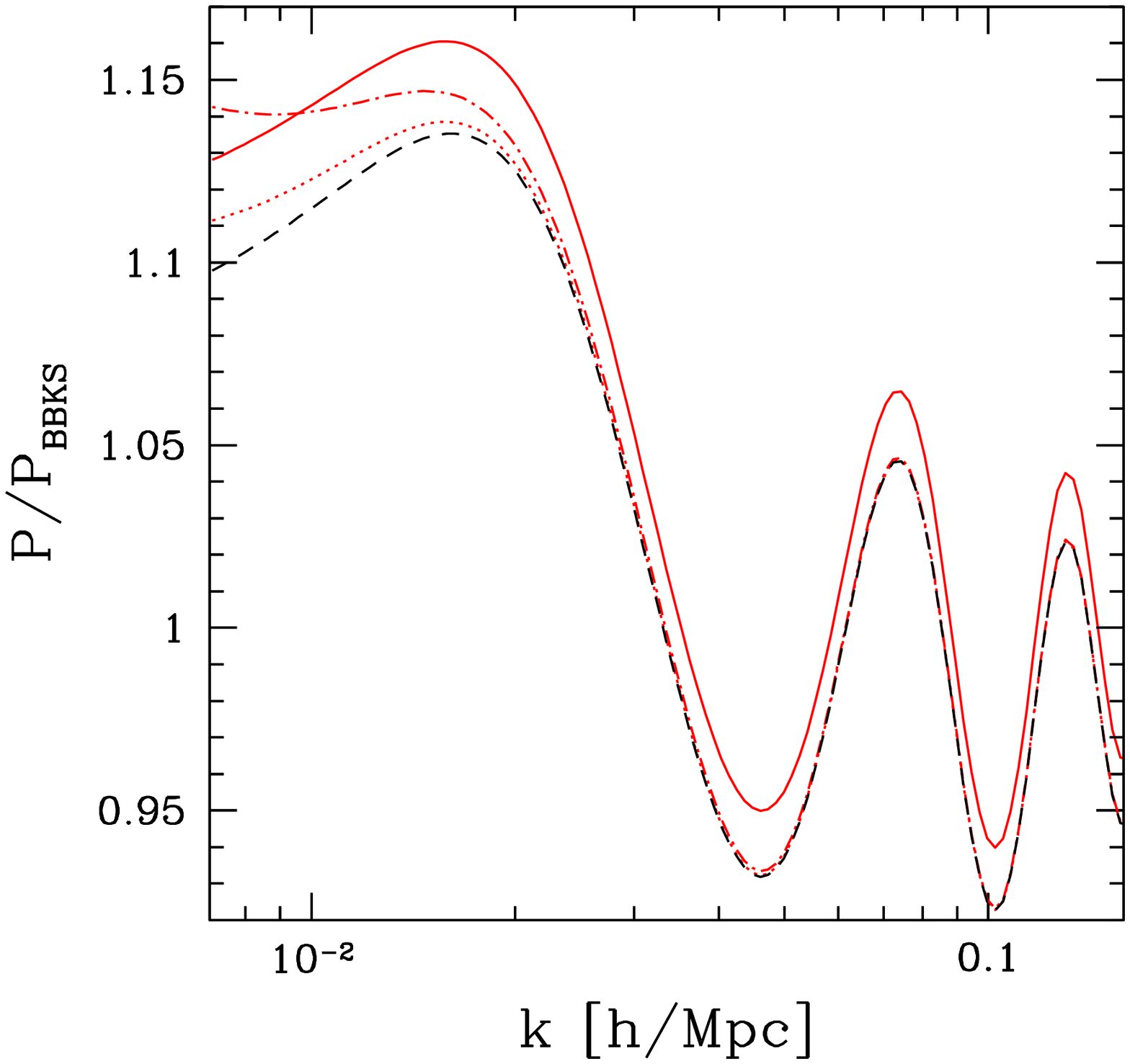}}
\caption{$\bar z = 1$: ({\bf a}) Contours of constant $P_{\rm obs}$ (solid) and
$P_{gg}$ (dotted), left to right: ${\rm log_{10} \,} (P/[{\rm Mpc/h}]^3) =$ 4.02 (red; double
contours), 3.93 (blue: double contours), 
3.83, 3.73, ... 3.33 (black). ({\bf b}) Various power spectra normalized by
the same BBKS (no baryon) galaxy power spectrum: $P_{\rm obs}$ for $k_\parallel = 0$
(red solid), $P_{gg}$ (black dashed), monopole of $P_{\rm obs}$ ($(5s-2)/b = 1$ for red dotted
and $(5s-2)/b = 2$ for red dot-dashed). 
Note that $k^2 = k_\parallel^2 + k_\perp^2$. 
A Gaussian window is assumed with
$\sigma = 204$ Mpc/h for both panels, and $(5s-2)/b = 1$ is adopted 
throughout except for the red dot-dashed curve.
}
\label{pkall1}
\end{figure*}

\begin{figure*}[tb]
\subfigure[]
{\label{pkcontour.new.try.1.5}\includegraphics[width=.45\textwidth]{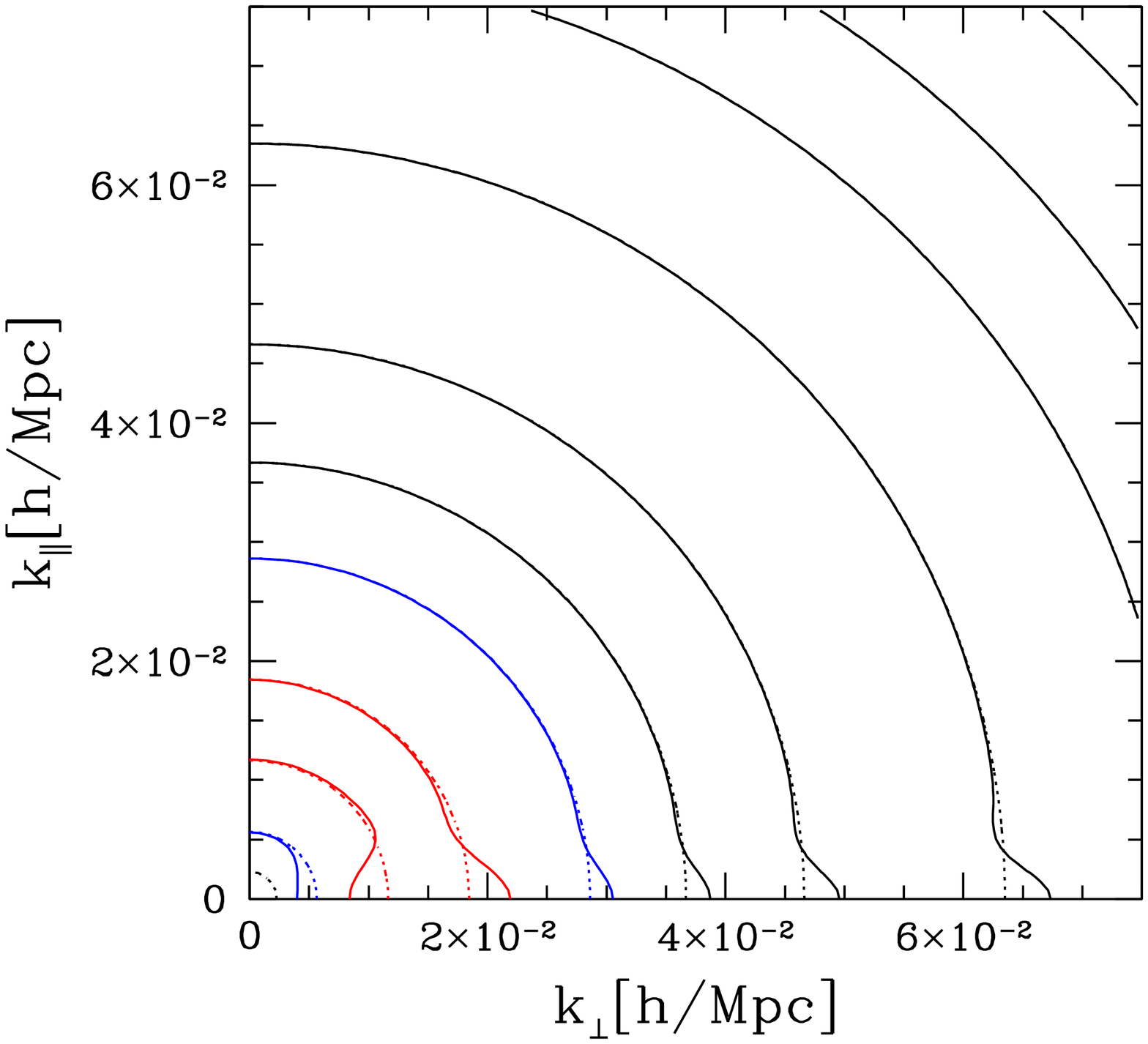}}
\hspace{0.1in}
\subfigure[]
{\label{pkBBKS2.trye2.1.5}\includegraphics[width=.45\textwidth]{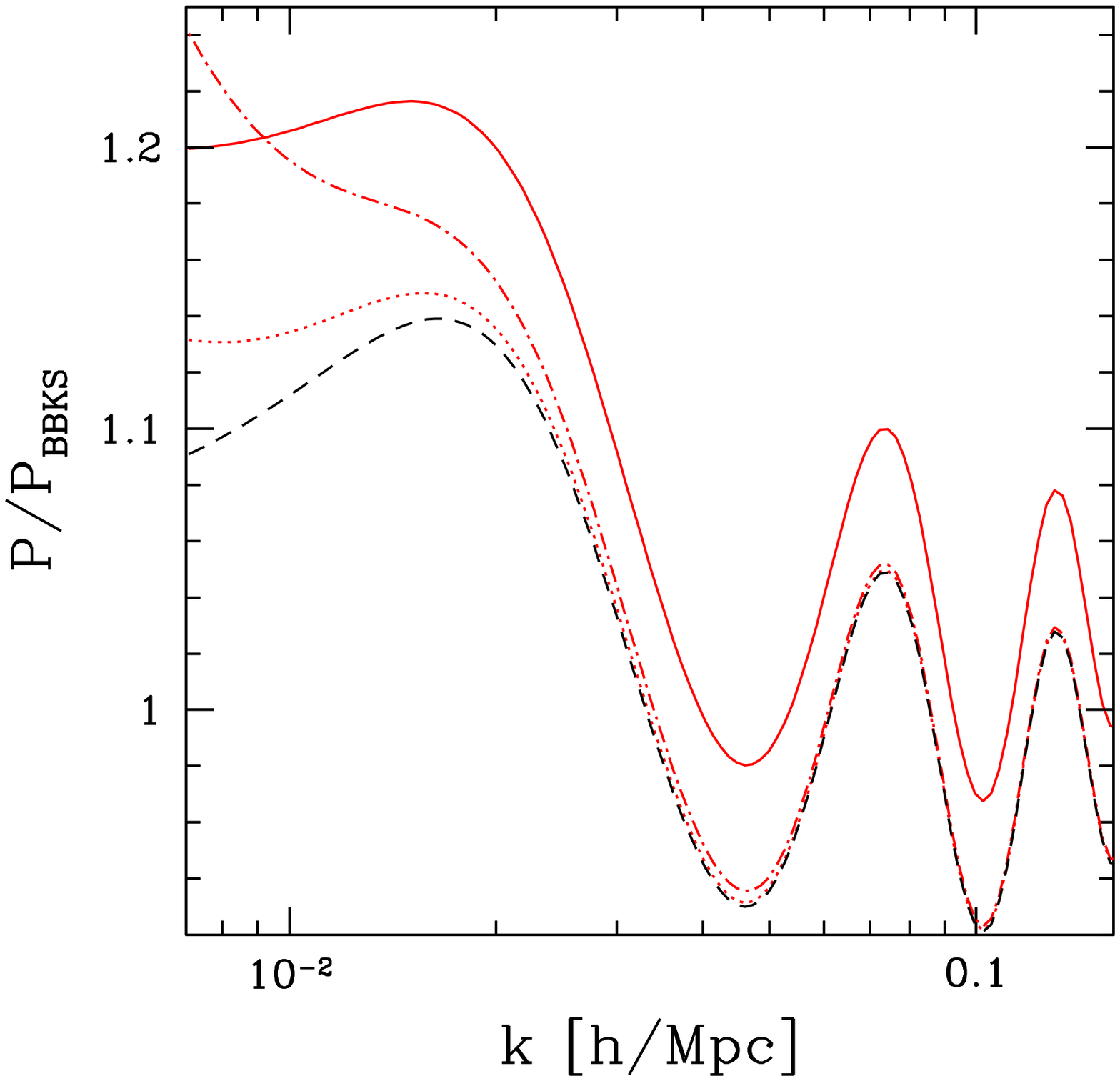}}
\caption{Analog of Fig. \ref{pkall1} for $\bar z= 1.5$. The contours in ({a}) are, left to right: 
${\rm log_{10} \,} (P/[{\rm Mpc/h}]^3) =$ 3.86 (red; double
contours), 3.76 (blue: double contours), 
3.66, 3.56, ... 3.16 (black). A Gaussian window is assumed with
$\sigma = 270$ Mpc/h.}
\label{pkall1.5}
\end{figure*}

\begin{figure*}[tb]
\subfigure[]
{\label{pkcontour.new.try.2}\includegraphics[width=.45\textwidth]{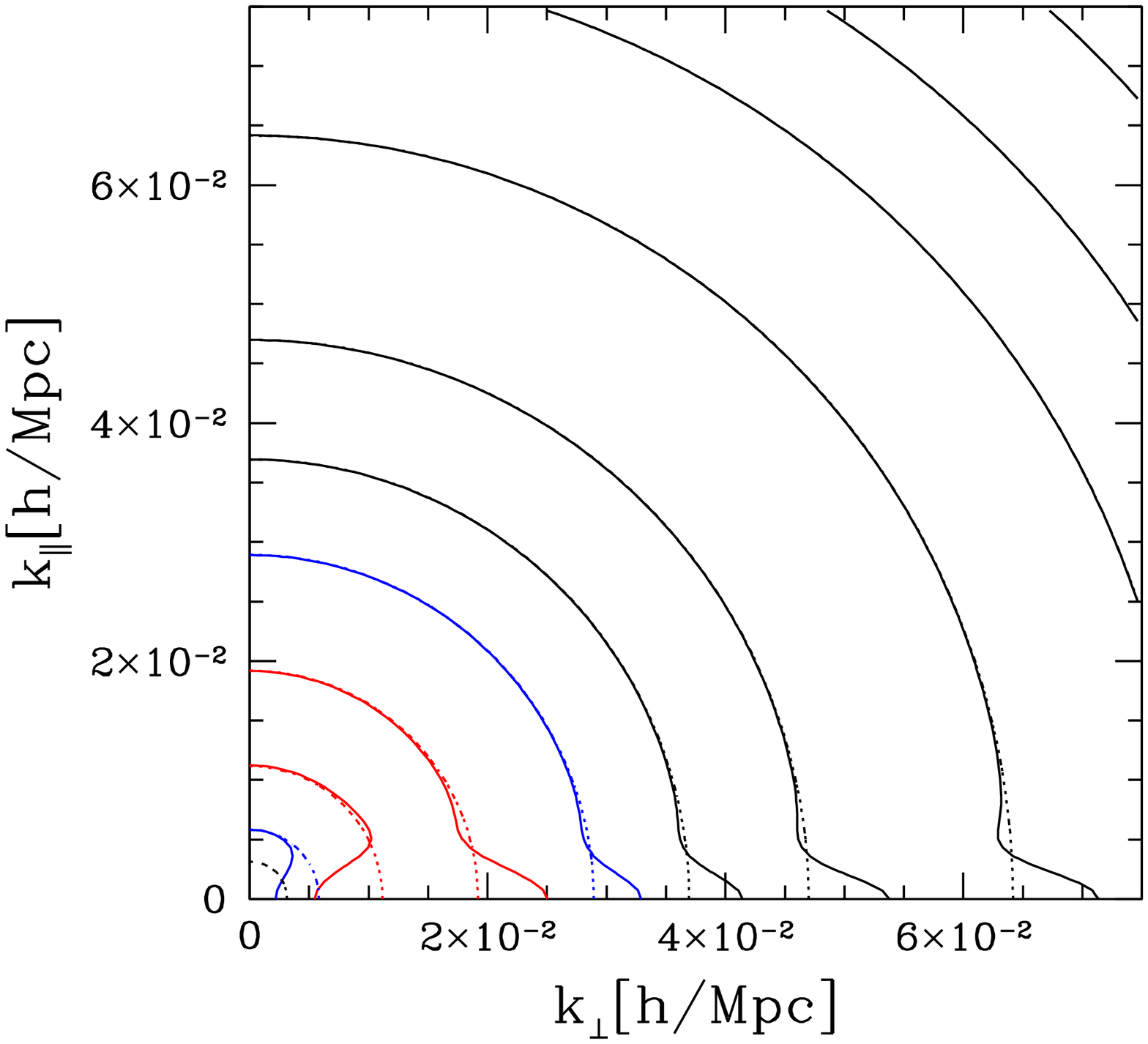}}
\hspace{0.1in}
\subfigure[]
{\label{pkBBKS2.trye2.2}\includegraphics[width=.45\textwidth]{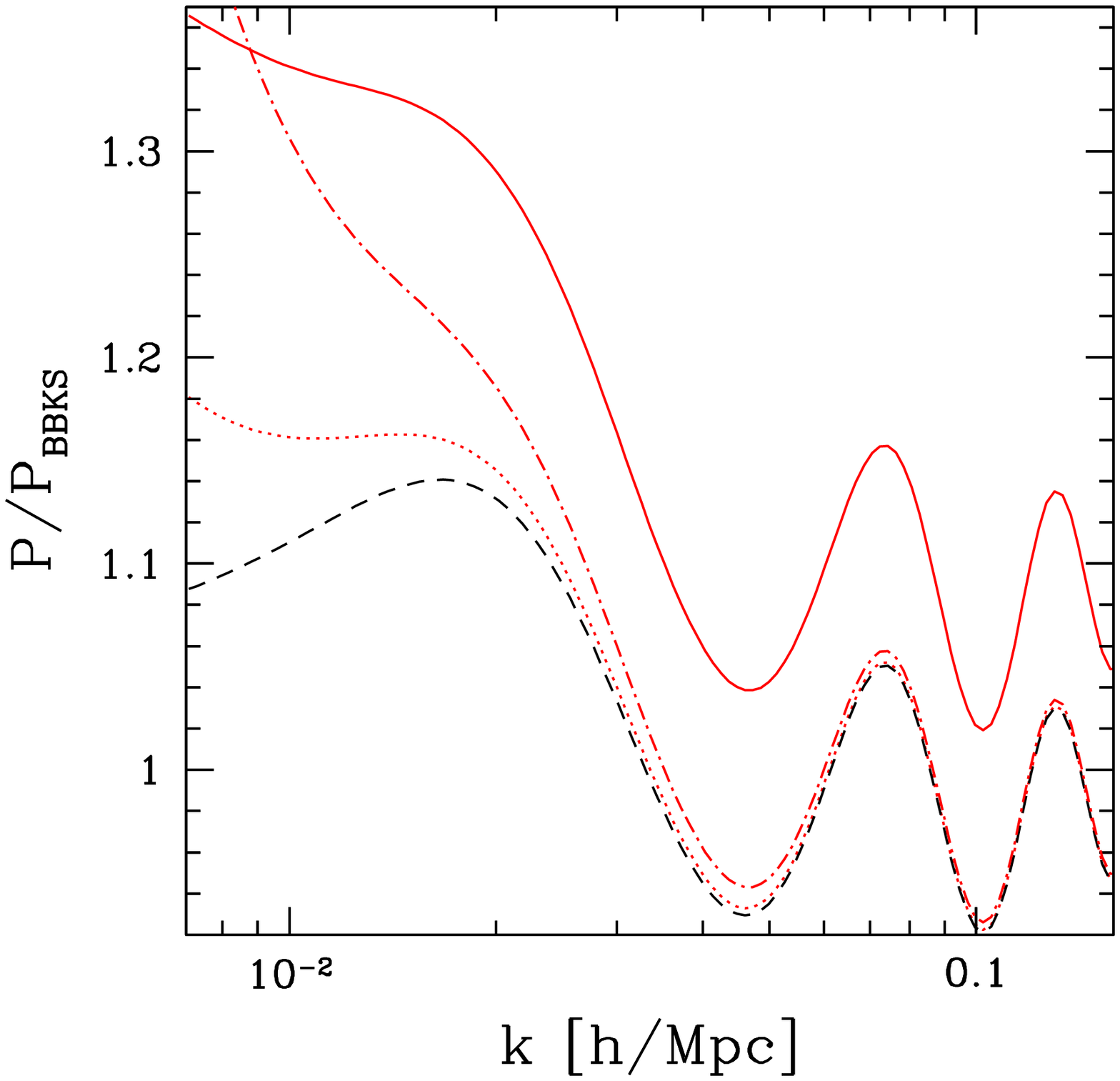}}
\caption{Analog of Fig. \ref{pkall1} for $\bar z = 2$. The contours in ({a}) are, left to right: 
${\rm log_{10} \,} (P/[{\rm Mpc/h}]^3) =$ 3.71 (red; double
contours), 3.61 (blue: double contours), 
3.51, 3.41, ... 3.01 (black). A Gaussian window is assumed with
$\sigma = 323$ Mpc/h.}
\label{pkall2}
\end{figure*}

\begin{figure*}[tb]
\subfigure[]
{\label{pkcontour.new.try.3}\includegraphics[width=.45\textwidth]{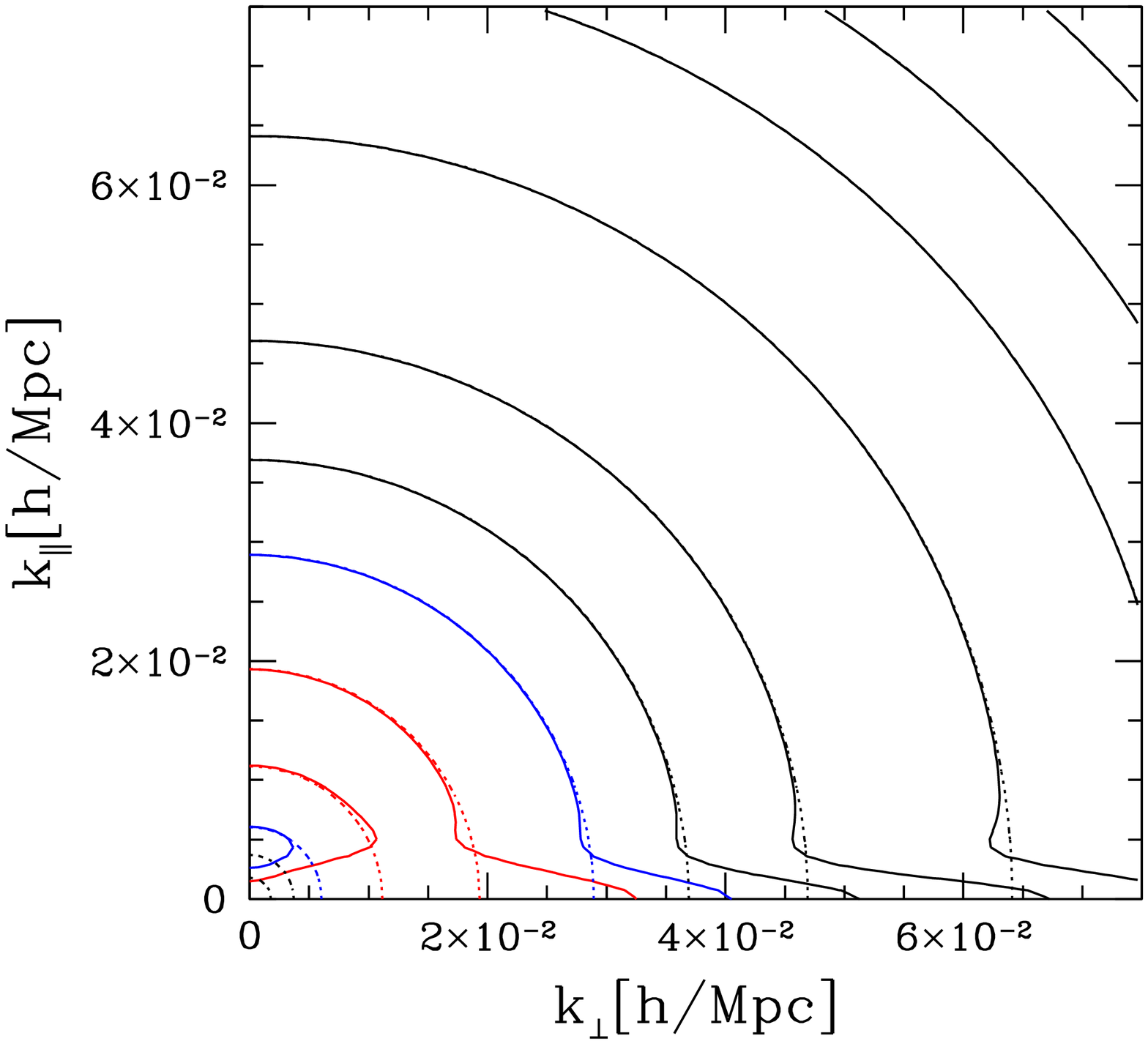}}
\hspace{0.1in}
\subfigure[]
{\label{pkBBKS2.trye2.3}\includegraphics[width=.45\textwidth]{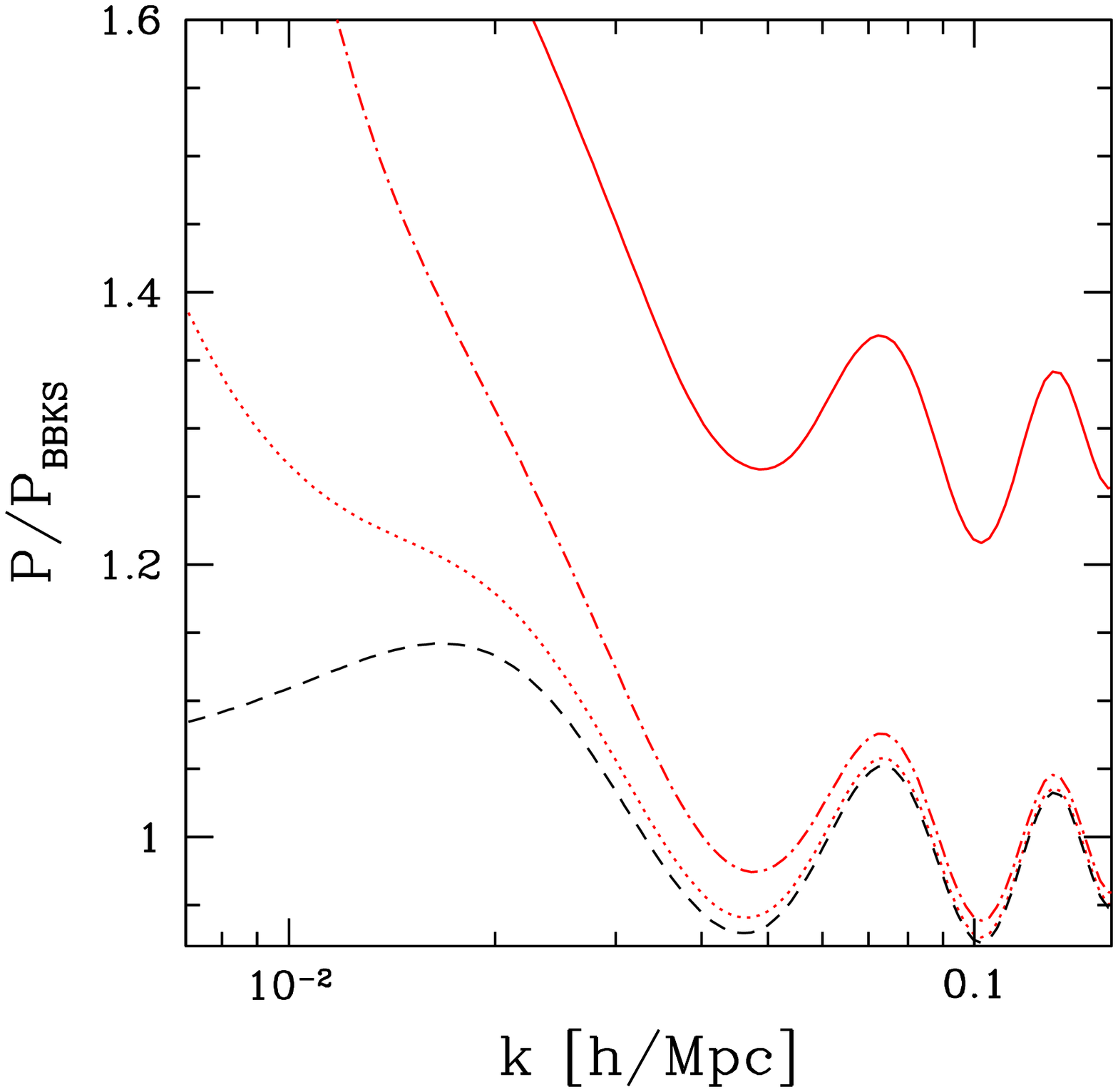}}
\caption{Analog of Fig. \ref{pkall1} for $\bar z = 3$. The contours in ({a}) are, left to right: 
${\rm log_{10} \,} (P/[{\rm Mpc/h}]^3) =$ 3.47 (red; double
contours), 3.37 (blue: double contours), 
3.27, 3.17, ... 2.77 (black). A Gaussian window is assumed with
$\sigma = 397$ Mpc/h.}
\label{pkall3}
\end{figure*}


For completeness, let us also give the window functions in the transverse directions for
a top-hat (a circle of radius $R$):
\begin{eqnarray}
W_\perp (x_\perp) &=& 1/\sqrt{\pi R^2} \quad {\rm if} \quad x_\perp < R \\ \nonumber
&=& 0 \quad {\rm otherwise} \\ \nonumber
\tilde W_\perp (k_\perp) &=& \left[{4\pi \over k_\perp^4 R^2}\right]^{1/2} \int_0^{k_\perp R} dr r J_0(r) \\ \nonumber
&=& {\sqrt{4\pi} \over k_\perp} J_1 (k_\perp R)
\end{eqnarray}
where $J_0$ and $J_1$ are the Bessel functions, 
and for a Gaussian:
\begin{eqnarray}
W_\perp (x_\perp) = (\pi\sigma^2)^{-1/2} {\,\rm exp\,} [-x_\perp^2/(2\sigma^2)] \\ \nonumber
\tilde W_\perp (k_\perp) = \sqrt{4\pi\sigma^2} {\, \rm exp\,} [-\sigma^2 k_\perp^2/2]
\end{eqnarray}

Note that in Fourier space the galaxy power spectrum $P_{gg}$ can itself be anisotropic if the
window function is anisotropic. 
In the Gaussian case, if the same $\sigma$ were chosen for both $\tilde W_\perp$ and
$\tilde W_\parallel$, the windowed galaxy power spectrum will remain isotropic.
In our computations below in this section, we adopt this special choice in order to
more clearly show the anisotropy induced by magnification bias. 
We have checked that using the top-hat geometry yields a rather similar 
lensing anisotropy as long as one makes the choice
$2R \sim L \sim \sqrt{12} \sigma$
(the latter equality is chosen such that the Gaussian $G$ and the top-hat $G$ have the same
low $k_\parallel$ limit). 

As in the case of the correlation functions, it
is useful to give order of magnitude estimates for the ratios
$P_{\mu\mu}/P_{gg}$ and $P_{g\mu}/P_{gg}$:
\begin{eqnarray}
\label{Porder}
{P_{\mu\mu} \over P_{gg}} &\sim& \left[5s-2 \over b\right]^2 {(1+ \bar z)^2 \over 50} (\bar\chi H_0)^3 
{|\tilde W_\parallel (k_\parallel) |^2 \over H_0^{-1}} {P_{mm}(k_\perp) \over P_{mm} (k)} \nonumber \\
{2P_{g\mu} \over P_{gg}} &\sim& \left[5s-2 \over b\right] {1+\bar z \over 2}
H_0^2 G(k_\parallel) {P_{mm}(k_\perp) \over P_{mm} (k)}
\end{eqnarray}
where $k^2 = k_\perp^2 + k_\parallel^2$. 
The above expressions are approximate. 
For instance,
we have approximated $P_{mm} (k_\perp \bar\chi/\chi')$ (eq. [\ref{Pmumu}]) by
$P_{mm} (k_\perp)$. 
Nonetheless, they agree with the exact numerical integration
to within factor of a few, and they illustrate several important points.

The presence of $|\tilde W_\parallel |^2 $ and $G$ means
that the effects of magnification bias are largest for low $k_\parallel$'s.
The simplest limit to consider is the one with
$k_\perp \gg k_\parallel$ (i.e. a ${\bf k}$ vector that is oriented
{\it transverse} to the LOS)
in which case the factors of $P_{mm}$ cancel out in the ratios,
since $k \sim k_\perp$. 
In the small $k_\parallel$ limit, 
$|\tilde W_\parallel|^2 \sim L$ and $G \sim L^2/3$, where
$L$ is the width of the redshift bin over which one is measuring
the power spectrum.
One can see that for instance at $\bar z \sim 1.5$ where $\bar\chi H_0 \sim 1$,
and for $(5s-2)/b \sim 1$, we have
$P_{\mu\mu}/P_{gg} \sim 0.1 H_0 L$ and 
$2 P_{g\mu}/P_{gg} \sim 0.4 (H_0 L)^2$. 
A choice of $L \sim 900$ Mpc/h (which corresponds to the redshift
interval $1.5 \pm 0.35$) yields $P_{\mu\mu}/P_{gg} \sim 0.03$
and $2 P_{g\mu}/P_{gg} \sim 0.04$. In other words, the total effect 
of magnification bias is quite modest, $\sim 7 \%$ in this configuration. 
It should be kept in mind that (1) this estimate increases with
$L$, and (2) the $P_{\mu\mu}/P_{gg}$ ratio increases strongly with
redshift: the cubic dependence on $(\bar \chi H_0)$ and quadratic dependence
on $1+\bar z$ means this ratio rises rapidly beyond $\bar z \sim 1.5$. 

Nonetheless, it is perhaps a little surprising that magnification bias
appears to have a much more modest effect on the observed clustering
in Fourier space compared to real space,
where the corresponding ratios $\xi_{\mu\mu}/\xi_{gg}$ and
$2\xi_{g\mu}/\xi_{gg}$ can reach order unity or even higher in the LOS
orientation (eq. [\ref{xiorder}]; see paper I for details).
The fundamental
reason is the absence in Fourier space of 
this potentially large boost factor, 
$\Delta^2 (k_*)/\Delta^2 (k_{**})$ in eq. (\ref{xiorder}), that is
present for the correlation function.
Consider a separation vector $\delta{\bf x}$ that is oriented along
the LOS, which is the orientation that maximizes
the magnification bias effect. If $|\delta{\bf x}|$ is sufficiently large,
the galaxy-galaxy correlation $\xi_{gg}$ is essentially determined
by the power spectrum in the linear regime (low $k_{**} \sim 1/|\delta{\bf x}|$ in eq. [\ref{xiorder}])
and is quite weak, while the galaxy-magnification 
correlation $\xi_{g\mu}$ and the magnification-magnification correlation
$\xi_{\mu\mu}$ are sensitive to the power in the nonlinear regime
(high $k_{*} \sim 1/\delta x_\perp$ in eq. [\ref{xiorder}]) and can be appreciable.
In other words, in the LOS orientation in real space, one is 
comparing intrinsic galaxy fluctuations and lensing fluctuations on very different scales.

Consider, on the other hand, a vector ${\bf k}$ that points in
the transverse direction, which is the Fourier analog of a LOS $\delta {\bf x}$.
In this case, $k \sim k_\perp$ and factors of the mass power spectrum simply cancel out
of the ratios in eq. (\ref{Porder}). There is no boost coming from a ratio
of powers on very different scales, as in the case of the correlation function.
{\it In other words, the mixing of magnification bias corrections with
the intrinsic galaxy clustering term occurs in a very different manner in Fourier space 
than in real space.}

Ultimately, the correlation function and the power spectrum are
related by Fourier transform, and so should really contain
the same information. The important point to keep in mind, however,
is that for cosmological purposes, one often focuses on scales that are perceived to be linear: 
in the case of the correlation function, that means a large $|\delta{\bf x}|$,
and in the case of the power spectrum, that means a small $|{\bf k}|$. 
With the presence of magnification bias, a large $|\delta{\bf x}|$ no longer
protects one against nonlinearity i.e. a separation $\delta{\bf x}$ pointing along the
LOS (i.e. small $|\delta {\bf x_\perp}|$)
is subject to large magnification corrections even if $|\delta{\bf x}|$ is large.
On the other hand, a small $|{\bf k}|$ means both $k_\parallel$ and $|{\bf k_\perp}|$
must be small, which protects one against nonlinear fluctuations. 

Let us return to the order of magnitude estimates in
eq. (\ref{Porder}), to see  what happens if one considers angular
averages. The monopole of any power spectrum $P$ is defined as
\begin{eqnarray}
\label{monopoleP}
{\rm monopole \,\, of \,\,} P({\bf k}) = \int_0^{\pi/2} P ({\bf k}) 
{\,\rm sin}\theta_k d\theta_k \, .
\end{eqnarray}
where $\theta_k$ is the angle between the LOS and ${\bf k}$. 
Let us focus on the monopole $P_{\mu\mu}/P_{gg}$ for a Gaussian
window, where the calculation is the simplest. 
Suppose one is interested in a scale $k$ such that $k\sigma \gsim 3$.
Because of the Gaussian in $|\tilde W_\parallel |^2$, the integral
over angle will be dominated by $\theta_k \sim \pi/2$, and so
one can approximate $P_{mm} (k_\perp)$ by $P_{mm} (k)$. The remaining
integral over angle is simple to do, and yields the ratio:
\begin{eqnarray}
\label{Pmonopoleratio}
{{\rm monopole \,\, of \,\,} P_{\mu\mu} \over {\rm monopole \,\, of \,\,} P_{gg}}
\sim \left[5s-2 \over b\right]^2 {(1+ \bar z)^2 \over 50} (\bar\chi H_0)^3 {\pi H_0 \over k}
\end{eqnarray}
which is valid {\it only} for $k\sigma \gsim 3$.
Note how the width of the window $\sigma$ completely disappears
from this ratio. A similar expression holds for the tophat case
as well. This ratio is small for $k/H_0 \gg 1$, unless one goes to
a sufficiently high redshift. 
The corresponding ratio for galaxy-magnification cannot
be worked out analytically because the window $G$ has a more complicated form.
On dimensional grounds, one expects this ratio to scale with $\sigma$.
In practice, we find that for redshifts where magnification bias matters,
the monopole of the galaxy-magnification power spectrum is quite a bit smaller than
that of the magnification-magnification power spectrum, in part because of cancellations that
occur under the angular average.

The intuition gained above from the order of magnitude estimates 
(eq. [\ref{Porder}], [\ref{Pmonopoleratio}]) is confirmed
by the exact numerical evaluation of $P_{g\mu}$, $P_{\mu\mu}$ and $P_{gg}$
according to eq. (\ref{Pgmu}), (\ref{Pmumu}) and (\ref{Pgg}). This is shown
in Fig. \ref{pkall1} - \ref{pkall3} for redshifts $\bar z = 1, 1.5, 2, 3$. The magnification distortion increases with redshift and, as discussed before, has the most noticeable effects for 
small $k_\parallel$'s. Exactly how small $k_{||}$ needs to be to see a substantial effect depends on the LOS width of the survey/sample selection function. For a Gaussian window function the region of large magnification distortion is $k_{||} \lsim \sigma^{-1}$. Panels (b) of Fig. \ref{pkall1} - \ref{pkall3} show that the monopole of the power spectrum is visibly distorted as well, the effect at low $k$'s is particularly severe at redshifts $\gsim 2$.

There is the interesting question of how magnification bias impacts
baryon oscillation measurements.
This was addressed in paper I for the real space correlation function.
{\it Briefly summarizing: we found that the observed baryon acoustic
scale can be shifted by up to $\sim 3 \%$ in the LOS orientation, and up
to $\sim 0.6 \%$ in the monopole, depending
on the exact values of the galaxy bias, redshift and number count slope.
The corresponding shifts in the inferred Hubble parameter and angular
diameter distance, if ignored, could significantly bias measurements of
the dark energy equation of state (by up to $\sim 15 \%$).} 
In Fourier space,
there are several wiggles, and the magnification bias induced shift
in the baryon oscillation scale is likely more sensitive to exactly
how this scale is extracted from data. We therefore do not attempt
to investigate this further in this paper.
Given the earlier discussions,
it is reasonable to expect that baryon acoustic oscillation measurements
are less affected by magnification bias in Fourier space.
However, it is worth emphasizing that magnification bias introduces
scale and orientation dependent corrections to the observed power spectrum,
and these corrections depend on uncertain factors such as the galaxy bias.
The question is whether, in fitting the observed data for the baryon
oscillation scale, one should introduce
additional fit parameters to account for magnification bias, and
what impact they might have on the measurement accuracy of the Hubble parameter
and the angular diameter distance. This certainly deserves more study.

It is also worth noting that, as can be seen from panels (b) of
Fig. \ref{pkall1} - \ref{pkall3}, the radiation-matter equality peak location/shape around
$k \sim 0.01$ h/Mpc is likely significantly affected by magnification bias,
and one must be careful in using it as a standard ruler \cite{asantha}.


\section{Incorporating Redshift Distortion due to Peculiar Motions}
\label{anisotropy}

\begin{figure*}[tb]
\subfigure[]
{\label{pkcontour.new.Z.sig3.0.2}\includegraphics[width=.45\textwidth]{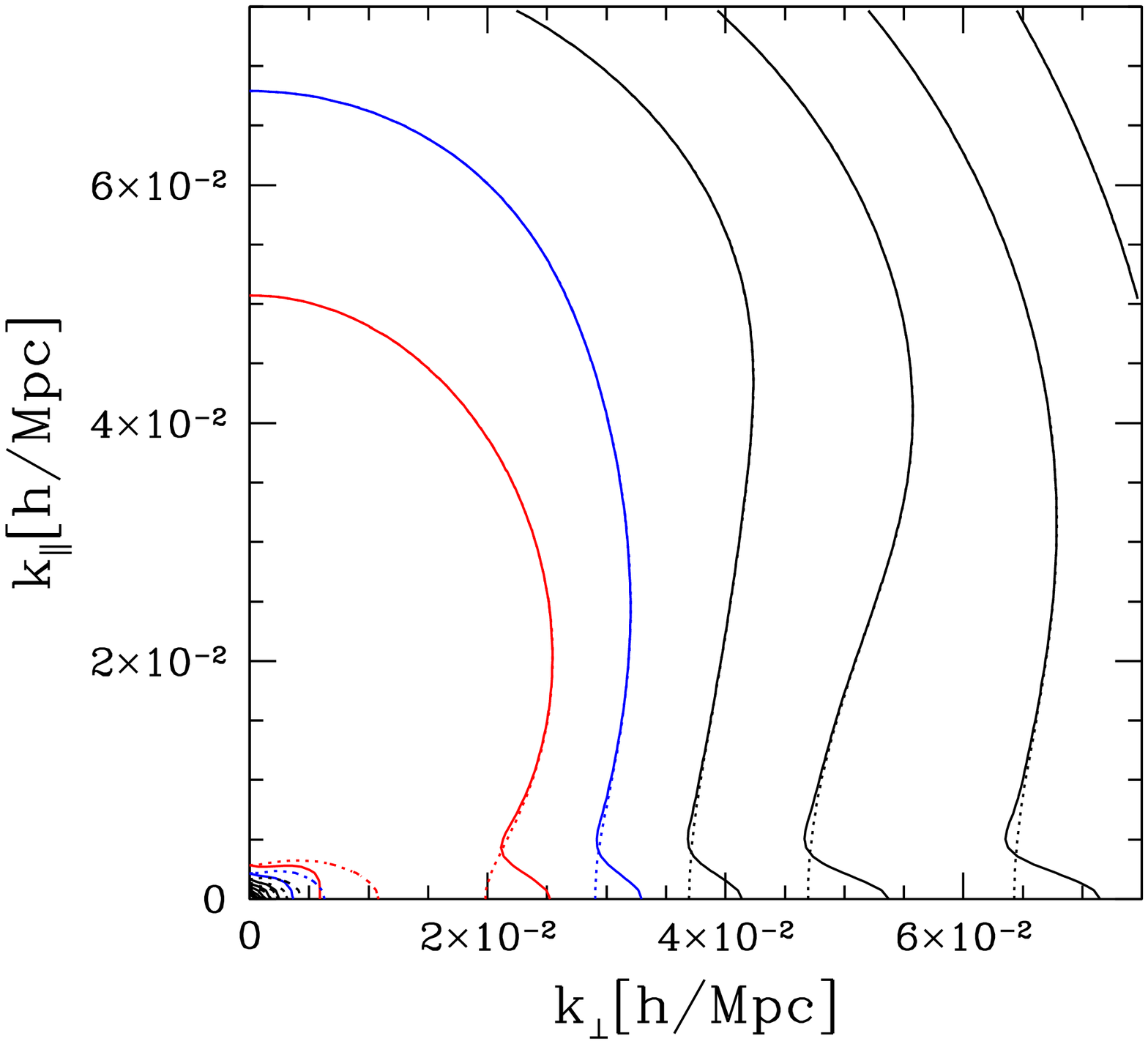}}
\hspace{0.1in}
\subfigure[]
{\label{pkBBKS2.trye2.Z.sig3.0.2}\includegraphics[width=.45\textwidth]{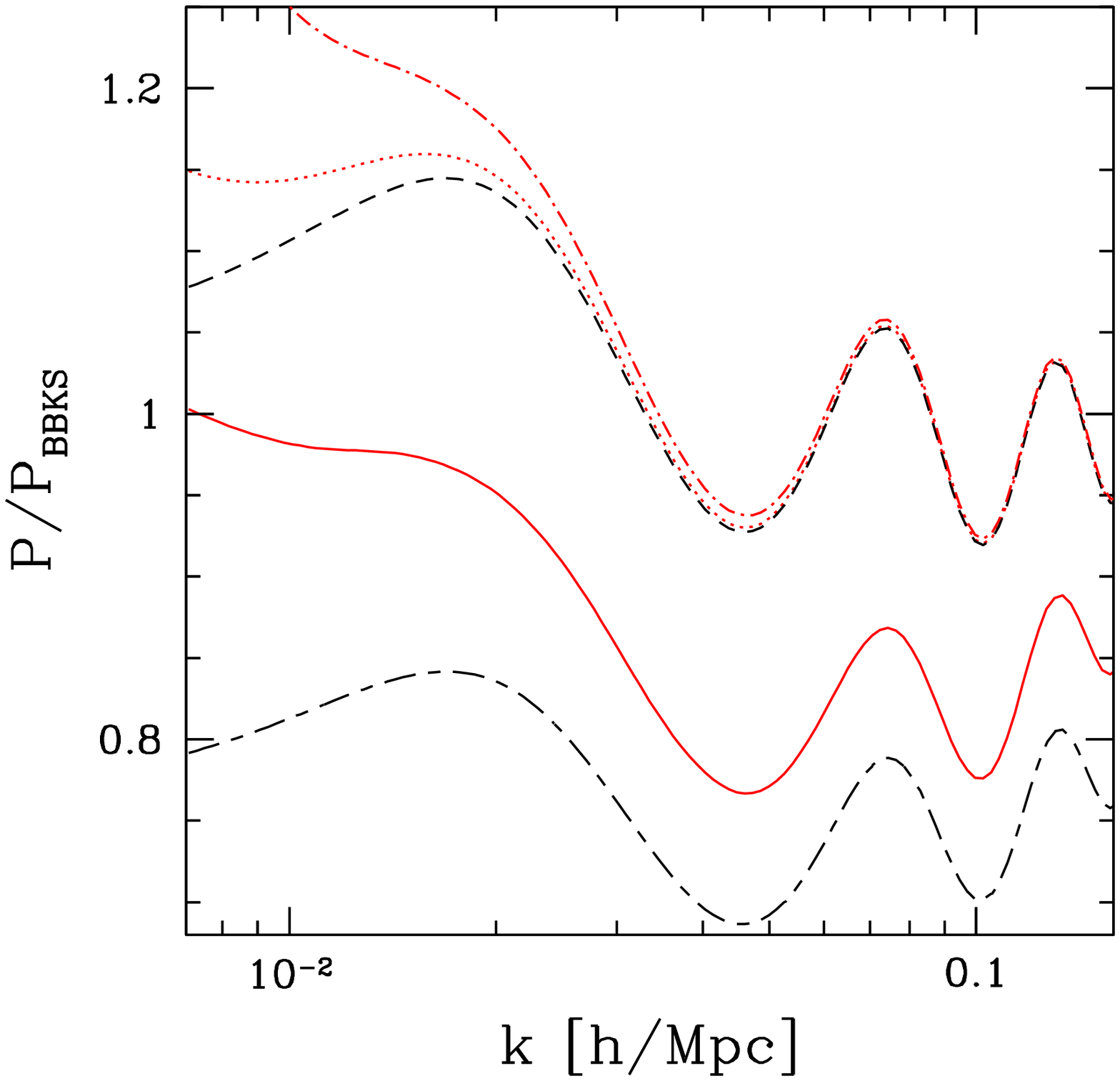}}
\caption{Analog of Fig. \ref{pkall2} for $\bar z = 2$, except 
redshift distortion is incorporated with $f_D/b = 0.475$, and $\sigma_z = 3$ Mpc/h
(eq. [\ref{approx}]).
({\bf a}) Contours of constant $P_{\rm obs}$ with magnification bias (solid)
and without magnification bias (dotted),
left to right: ${\rm log_{10} \,} (P/[{\rm Mpc/h}]^3) =$ 3.71 (red; double
contours), 3.61 (blue: double contours), 
3.51, 3.41, 3.31, 3.21 (black). 
({\bf b}) Various power spectra normalized by
the same BBKS (no baryon) galaxy monopole power spectrum: $P_{\rm obs}$ with magnification bias 
for $k_\parallel = 0$
(red solid), $P_{\rm obs}$ without magnification bias for $k_\parallel = 0$
(black short-long dashed), monopole of $P_{\rm obs}$ without magnification bias
(black uniform dashed), monopole of $P_{\rm obs}$ with magnification bias ($(5s-2)/b = 1$ for red dotted
and $(5s-2)/b = 2$ for red dot-dashed). 
Note that $k^2 = k_\parallel^2 + k_\perp^2$. 
A Gaussian window is assumed with
$\sigma = 323$ Mpc/h for both panels, and $(5s-2)/b = 1$ is adopted 
throughout except for the red dot-dashed curve.
}
\label{pkall.Z.3.0.2}
\end{figure*}

\begin{figure*}[tb]
\subfigure[]
{\label{pkcontour.new.Z.2}\includegraphics[width=.45\textwidth]{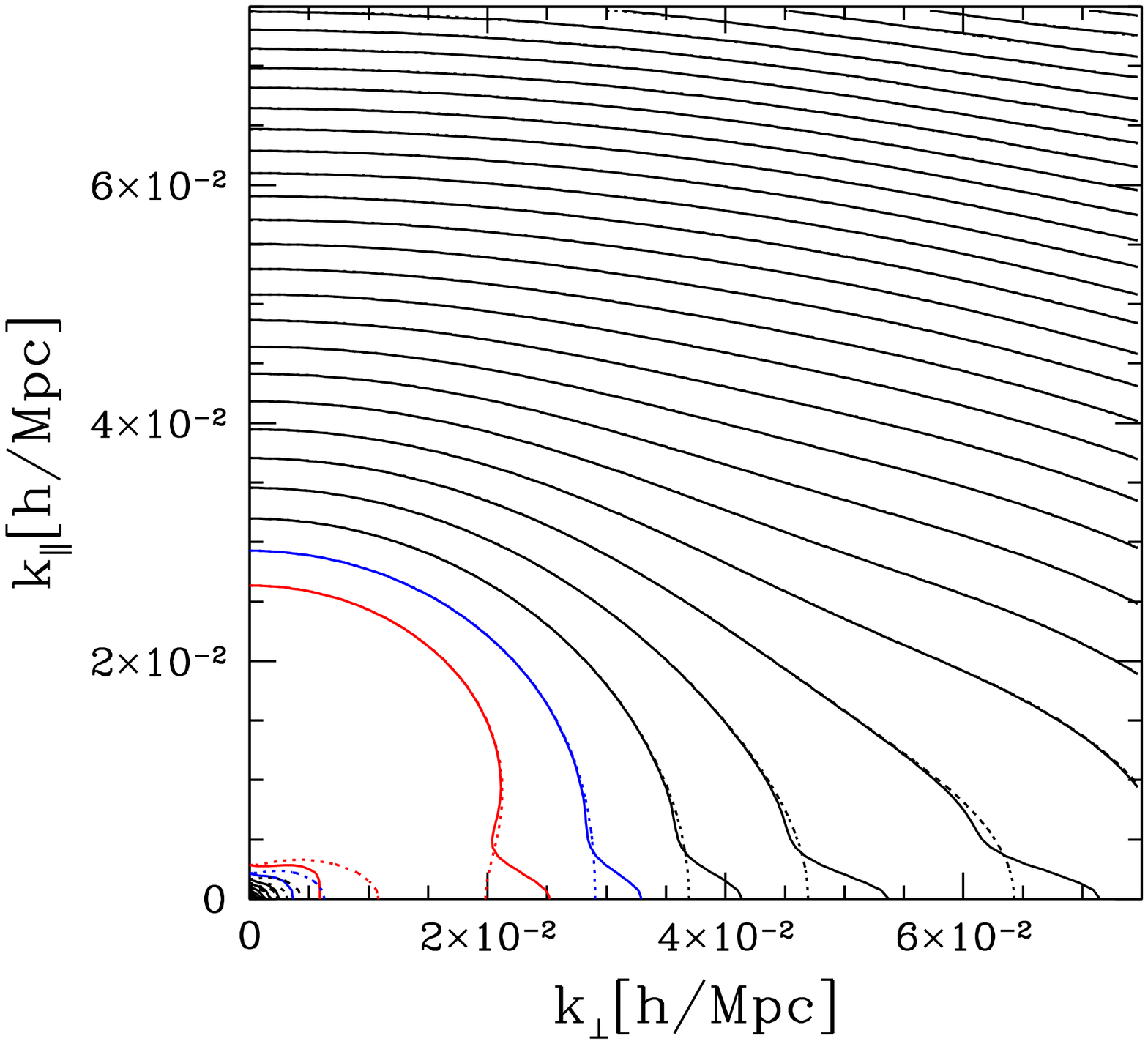}}
\hspace{0.1in}
\subfigure[]
{\label{pkBBKS2.trye2.Z.2}\includegraphics[width=.45\textwidth]{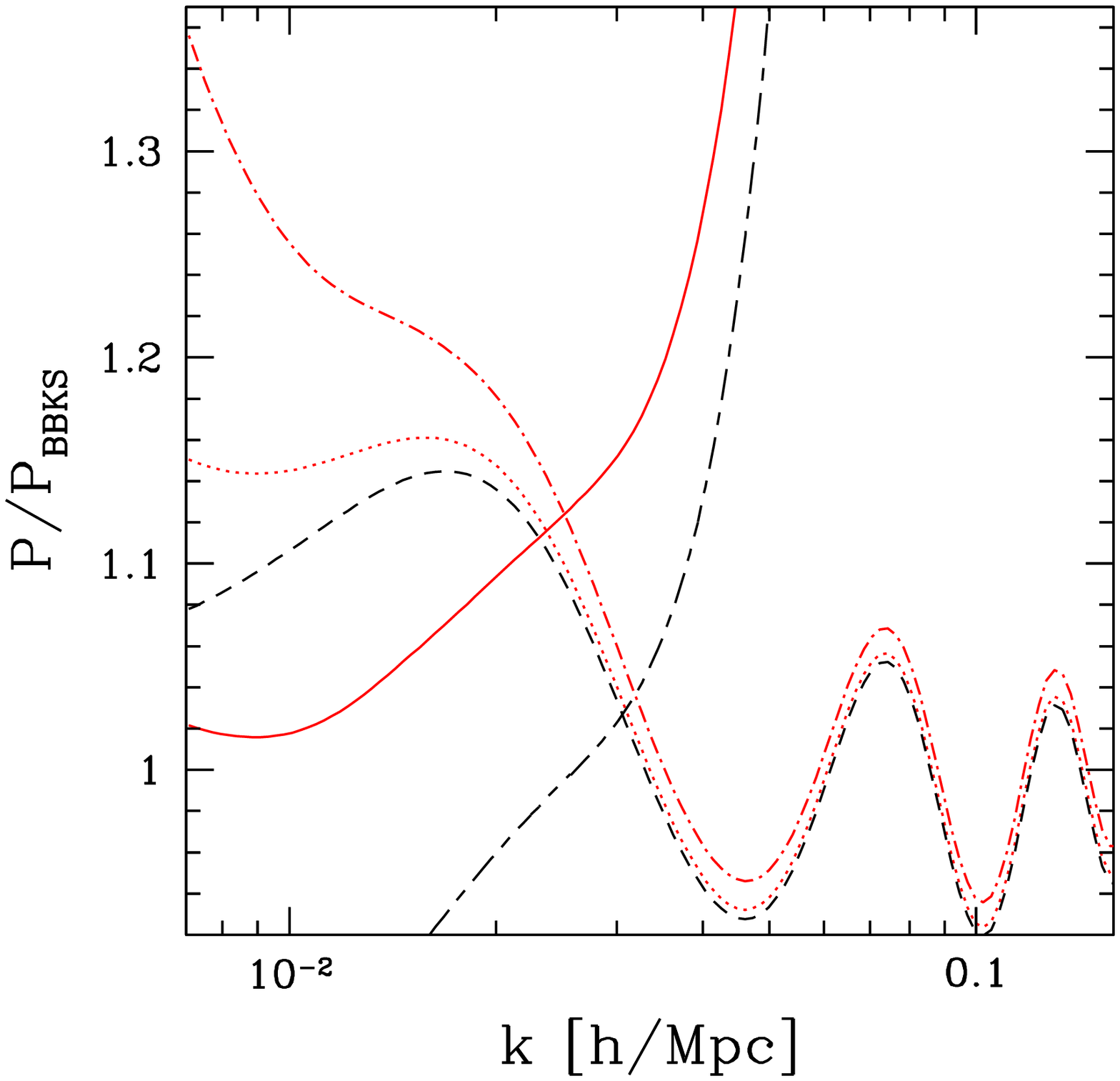}}
\caption{Analog of Fig. \ref{pkall.Z.3.0.2} except that $\sigma_z = 30$ Mpc/h. 
The contours in (a) are, left to right: ${\rm log_{10} \,} (P/[{\rm Mpc/h}]^3) =$ 3.71 (red; double
contours), 3.61 (blue: double contours), 
3.51, 3.41, ... 1.01 (black). In panel (b), note how the monopole
(red dotted  with magnification bias
and black uniform dashed without) is higher than
the $k_\parallel = 0$ power (red solid with magnification bias
and black short-long dashed without) for small $k$'s
due to the Kaiser effect, and lower for high $k$'s due to
the finger-of-god effect (from the large $\sigma_z$).
}
\label{pkall.Z.2}
\end{figure*}
The observed redshift of a source galaxy is dependent upon both the distance to the source (the cosmological redshift) and the peculiar velocity of the source.  In redshift space, the observed galaxy density is (to first order in perturbations):
\begin{eqnarray}
\label{zspace}
\delta_{\rm obs} = \delta_g + \delta_\mu + \delta_v
\end{eqnarray}
where $\delta_g$ and $\delta_\mu$ are as in eq. (\ref{start}), and $\delta_v$ is
\begin{eqnarray}
\delta_v = - {(1+\bar z) \over H(\bar z)} {\partial v_\parallel \over \partial x_\parallel}
\end{eqnarray}
where $v_\parallel$ is the LOS peculiar velocity, 
and $H(\bar z)$ is the Hubble parameter at the mean redshift $\bar z$.

The observed two-point correlation function, instead of eq. (\ref{obs}), is now
given by
\begin{eqnarray}
\label{obsv}
&& \xi_{\rm obs} (1; 2) 
= \xi_{gg} (1; 2) + \xi_{g\mu} (1; 2) 
+ \xi_{g\mu} (2; 1) \\ \nonumber 
&& \quad + \xi_{\mu\mu} (1; 2) + \xi_{gv} (1;2) + \xi_{vg} (2;1) + \xi_{vv} (1;2)
\end{eqnarray}
where we have used the arguments $1$ and $2$ as the shorthand for the positions
of the two points of interest in redshift-space.
We have used the Limber approximation which makes the velocity-magnification cross-terms vanish
(the derivative with respect to $x_\parallel$ in the velocity term pulls down a factor of
$k_\parallel$, which vanishes under the Limber approximation; see \cite{matsubara00}).
The galaxy-velocity cross-correlation and the velocity auto-correlation are 
given by the well-known results of Kaiser \cite{kaiser}:
\begin{eqnarray}
\xi_{gv} (1,2) = \left[ {a \over a'} {D' \over D} \right]
\int {d^3 k \over (2\pi)^3} {k_\parallel^2 \over k^2} P_{gm} (k) e^{i {\bf k} \cdot ({\bf x_1} - {\bf x_2})}
\end{eqnarray}
and
\begin{eqnarray}
\xi_{vv} (1,2) = \left[ {a \over a'} {D' \over D} \right]^2
\int {d^3 k \over (2\pi)^3} {k_\parallel^4 \over k^4} P_{mm} (k) e^{i {\bf k} \cdot ({\bf x_1} - {\bf x_2})}
\end{eqnarray}
where ${\bf x_1} - {\bf x_2}$ has a LOS component $\chi_1 - \chi_2$, and 
transverse components of $\bar\chi (\thetaB_1 - \thetaB_2)$. Here, $a$ is the scale factor,
$D$ is the linear growth factor, and
$a'$ and $D'$ are their derivatives with respect to conformal time. All time dependent quantities
are evaluated at the mean redshift $\bar z$.

The observed power spectrum, instead of eq. (\ref{Pobs}), is now given by
\begin{eqnarray}
&& P_{\rm obs} ({\bf k}) = P_{gg} ({\bf k}) + 2 P_{g\mu} ({\bf k}) + P_{\mu\mu} ({\bf k}) \\ \nonumber 
&& \quad \quad +
2 P_{gv} ({\bf k}) + P_{vv} ({\bf k})
\end{eqnarray}
where the first line is as before (eq. [\ref{Pgmu}], [\ref{Pmumu}] \& [\ref{Pgg}]), and 
\begin{eqnarray}
P_{gv} ({\bf k}) = \left[ {a \over a'}{D' \over D} \right]
\int {d^3 k' \over (2\pi)^3} {{k_\parallel'}^2\over {k'}^2} P_{gm} (k') |\tilde W({\bf k} - {\bf k'})|^2
\end{eqnarray}
and 
\begin{eqnarray}
P_{vv} ({\bf k}) = \left[ {a \over a'}{D' \over D} \right]^2
\int {d^3 k' \over (2\pi)^3} {{k_\parallel'}^4 \over {k'}^4} P_{mm} (k') |\tilde W({\bf k} - {\bf k'})|^2
\end{eqnarray}

To gain some intuition about the various effects at work, it is useful to adopt
the following approximation: integrate out all the convolving windows as if
they are delta functions.
(This approximation is made only in this section, {\it not} in previous sections.)
For instance:
\begin{eqnarray}
\label{Pwinapprox}
\int {d^3 k' \over (2\pi)^3} P (k') |\tilde W({\bf k - \bf k'})|^2
\sim P (k) 
\end{eqnarray}
where $P$ represents $P_{gg}$, $P_{gm}$, $P_{gm} {k'_\parallel}^2
/{k'}^2$ and so on, and the convolving window
$\tilde W$ could also be $\tilde W_\perp$.
This works well if the power spectrum is sufficiently smooth and the convolving
window is sufficiently narrow. With this approximation, we obtain
\begin{eqnarray}
\label{approx}
&& P_{\rm obs} (\bar z, {\bf k}) = P_{gg} (\bar z, k) 
\Bigl[ \left(1 + {f_D\over b}{k_\parallel^2 \over k^2}\right)^2 e^{-k_\parallel^2 \sigma_z^2} + \\ \nonumber
&& q (1+\bar z) G(k_\parallel) {P_{mm}(\bar z, k_\perp) \over 
  P_{mm}(\bar z, k)} + \\ \nonumber
&& q^2 |\tilde W_\parallel (k_\parallel) |^2 
   \int_0^{\bar\chi} d\chi' (\bar\chi - \chi')^2 (1+z')^2 {P_{mm}(z', k_\perp\bar\chi/\chi') \over
   P_{mm} (\bar z, k)} \Bigr]
\end{eqnarray}
where 
\begin{eqnarray}
\label{fDdef}
f_D \equiv { d {\,\rm ln\,} D\over d{\,\rm ln\,} a}
\end{eqnarray}
evaluated at the mean redshift $\bar z$, and
\begin{eqnarray}
\label{qdef}
q \equiv {3\over 2} H_0^2 {\Omega_m} {(5s-2) \over b}
\end{eqnarray}
and we have assumed a linear galaxy bias $b$ (at redshift $\bar z$).
We have introduced an exponential factor ${\,\rm exp}[-k_\parallel^2 \sigma_z^2]$
which accounts for a possible (Gaussian) dispersion in redshifts, such as from
photometric redshifts. Here $\sigma_z$ is the
dispersion expressed in comoving Mpc/h (not in redshift). 
One can also think of this exponential factor as modeling
the effect of nonlinear or virialized peculiar motions, though this description is at
best approximate \cite{roman}. Strictly speaking, with a non-zero $\sigma_z$, the
multiplicative window $G$ in eq. (\ref{approx}) should be replaced by
\begin{eqnarray}
\label{Gdef2}
&& G_z (k_\parallel) = \int {dz_1 dz_2 dz'_1 dz'_2 \over 2\pi \sigma_z^2}
|z'_1 - z'_2| 
e^{ik_\parallel (z_1 - z_2)} \\ \nonumber 
&& W_\parallel(z_1) W_\parallel (z_2)
{\,\rm exp}\left[-{(z'_1-z_1)^2\over 2 \sigma_z^2}\right]
{\,\rm exp}\left[-{(z'_2 - z_2)^2 \over 2 \sigma_z^2}\right]
\end{eqnarray}
where $z_1, z_2, z'_1, z'_2$ denotes LOS distances, not redshifts.
We find that as long as the LOS width of the survey/sample selection
$W_\parallel$ is large compared to $\sigma_z$
(i.e. $L \gg \sigma_z$ for a top-hat geometry, 
or $\sigma \gg \sigma_z$ for a Gaussian geometry; see eq.
[\ref{tophatFT}] \& [\ref{gaussFT}]), 
the above $G_z$ is well approximated
by $G$ as defined in eq. (\ref{Gdef}).

The first line of eq. (\ref{approx}), excluding the
exponential factor, represents the classic Kaiser distortion due to coherent infall.
The second and third lines come from the galaxy-magnification and 
magnification-magnification correlations respectively.
We illustrate all these effects in Fig. \ref{pkall.Z.3.0.2} and \ref{pkall.Z.2}
by evaluating eq. (\ref{approx}) 
for $\bar z = 2$. Fig. \ref{pkall.Z.3.0.2} uses $\sigma_z = 3$ Mpc/h, which corresponds to
a velocity dispersion of $\sim 300$ km/s, or a redshift dispersion of
$0.003$. This is the level of dispersion one expects from virialized motions
on small scales. 

For the scales shown in Fig. \ref{pkall.Z.3.0.2}, the Kaiser distortion
dominates over the finger-of-god effect due to virialized motions: 
the contours of constant $P_{\rm obs}$ are elongated in the LOS direction
for $k$'s larger than the radiation-matter equality scale, and compressed otherwise.
It is also worth noting that magnification distortion survives the Kaiser
distortion. The two kinds of distortions have fundamentally different shapes.
Magnification distortion is localized to small $k_\parallel$'s, whereas
the Kaiser distortion is more spread out. 
The localized nature of magnification distortion originates from the fact that
the corrections it introduces (the second and third terms on the right
hand side of eq. [\ref{approx}]) are proportional to 
the multiplicative windows $G(k_\parallel)$ and $| \tilde W_\parallel (k_\parallel) |^2$,
both of which peak at small $k_\parallel$'s.

Interestingly, the Kaiser distortion
vanishes at $k_\parallel = 0$, exactly where magnification bias has the largest effect.
This suggests it should be possible to disentangle the two different distortions 
from data.
For instance, the one free parameter that controls the Kaiser distortion, $f_D/b$,
can be determined from the observed power spectrum anisotropy by excluding from
consideration the small $k_\parallel$
modes. With this in hand, one should be able to predict the dotted contours such as those
in Fig. \ref{pkall.Z.3.0.2}a. The difference between the observed (solid) contours 
and the dotted ones then gives us the magnification bias corrections.
The galaxy-magnification and magnification-magnification contributions can be further
separated from each other by using the distinctive shapes of their respective
multiplicative windows $G$ and $|\tilde W_\parallel (k_\parallel)|^2$
(for instance, the latter is positive definite whereas the former can
go negative; see Fig. \ref{gwinoutT}).


Fig. \ref{pkall.Z.2} is analogous to Fig. \ref{pkall.Z.3.0.2} except that
$\sigma_z$ is increased
to $30$ Mpc/h, corresponding to a velocity dispersion of 
$\sim 3000$ km/s, or a redshift dispersion of $0.03$. This larger
value for $\sigma_z$ is chosen to 
mimic the effect of photometric redshifts. 
In contrast to the case of $\sigma_z = 3$ Mpc/h, one can clearly see here
a finger-of-god effect of sorts: the contours of constant $P_{\rm obs}$ are compressed
in the LOS direction, for sufficiently large $k$'s.
As before, magnification distortion is clearly visible, being
well localized to small $k_\parallel$'s. Its distinctive shape makes it in principle
distinguishable from both the Kaiser distortion
and the finger-of-god effect, which are more spread out on
the $k_\perp - k_\parallel$ plane.

It is also worth noting that in cases 
where the power spectrum is strongly anisotropic,
such as in Fig. \ref{pkall.Z.2},
the monopole is probably not the most relevant quantity to consider.
This is because some orientations are much more noisy than
the others, and one might not want to weigh them equally
(or more precisely, according to eq. [\ref{monopoleP}]).
In general, with sufficient redshift accuracy, one should 
make use of the full 3D information available. 
In this vein, it is not uncommon to consider higher multipoles
of the anisotropic power spectrum \cite{Hamilton}.
This is especially useful for analyzing the Kaiser distortion since
it gives rise to only 2 extra multipoles. 
However, a multipole expansion
is likely not helpful in analyzing magnification
distortion, due to its localized nature. It is
probably more useful to take advantage of the special
dependence on $G(k_\parallel)$ and 
$|\tilde W_\parallel (k_\parallel)|^2$ of
the galaxy-magnification and magnification-magnification
terms (eq. [\ref{approx}]). How to optimally extract these
two contributions from noisy data deserves further study.

In addition to redshift distortion due to peculiar motions,
another well known effect is the so called cosmological distortion,
or Alcock-Paczynski effect
\cite{AP,matsubara}. This is taken up in Appendix \ref{app:AP}.

\vspace{0.5cm}

\section{Discussion}
\label{discuss}

In paper I, we examined the effects of magnification distortion in real/configuration space,
and here we have extended the analysis to Fourier and redshift space.
The observed galaxy/quasar correlation is endowed with a distinctive lensing induced anisotropy.
This is encapsulated in eq. (\ref{scaling2}) for real space
and eq. (\ref{Pgmu}), (\ref{Pmumu}), (\ref{Pgg}) and (\ref{Pobs}) for Fourier space:
the linear dependence on the LOS separation
$\delta\chi$ in the real space galaxy-magnification correlation
gives rise to the multiplicative window $G$ in the galaxy-magnification power spectrum.
 Likewise, the independence of the real space magnification-magnification correlation
on $\delta\chi$ accounts for the appearance of the multiplicative window
$|\tilde W_\parallel |^2$ in the magnification-magnification power spectrum.

Qualitatively, the galaxy correlation becomes enhanced 
for the transverse modes in Fourier space, and for pairs oriented along the LOS in real space.
Quantitatively, the degree of enhancement is rather different for the dual spaces.
As explained in \S \ref{intro} and \S \ref{pk}, magnification distortion is
less severe in Fourier space: as long as one focuses on modes with
a small wavenumber $k$, as is usually done when obtaining cosmological constraints,
both the intrinsic galaxy fluctuations and the lensing
fluctuations are in the linear regime. 
In real space, even if the pair separation is large,
one cannot help but mix up linear intrinsic galaxy fluctuations
with nonlinear lensing fluctuations, as long as one considers the LOS orientation.
Incidentally, this implies that a linear galaxy bias is a better approximation in
Fourier space than in real space.

The above findings suggest that in precision measurements of the galaxy
power spectrum, such as those that attempt to use the baryon oscillation scale
to constrain dark energy \cite{eisenstein,2dFa,2dFb,hutsi,tegmark,baotheory,baoexp}, 
the simplest way to immunize against magnification bias could be to go to Fourier space,
and remove from consideration the small $k_\parallel$ (transverse) modes where magnification
bias has the largest effect.
However, in a photometric redshift survey, these are probably the
modes with the highest signal-to-noise, and the Fourier space also suffers
from possible complications due to a non-Poissonian shot noise \cite{sss}. 
A better strategy is perhaps to face the magnification bias corrections head on,
and use the full 3D information to constrain and measure them -- after all, they
contain interesting cosmological information too.

Precisely how magnification bias corrections might shift the baryon oscillation
scale is a subject worthy of a separate paper.
The precise shift will depend on exactly how the oscillation scale is extracted
from data. A preliminary investigation in real space, where the acoustic oscillations
manifest as a single local maximum whose position is easily defined, was presented
in paper I: as discussed at the end of \S \ref{pk}, the impact on measurements
of the dark energy equation of state can be significant (shifting it by up to
$\sim 15 \%$).
It should also be emphasized that the baryon oscillations are not
the only large scale features of interest. The radiation-matter equality
peak at around $k \sim 0.01$ h/Mpc contains valuable cosmological information,
but it can be seen from Fig. \ref{pkall1} - \ref{pkall3} panels (b) that
the power spectrum at this scale is likely significantly affected by magnification bias.

We have incorporated the effects of peculiar motion (and redshift inaccuracy) 
on the power spectrum anisotropy in \S \ref{anisotropy}
(the Alcock-Paczynski effect is further incorporated in Appendix \ref{app:AP}).
The main conclusion is that the lensing induced features remain rather robust,
thanks to their localized nature to small $k_\parallel$'s. 
It should be in principle possible to separately measure from data
the galaxy-galaxy, galaxy-magnification and magnification-magnification power spectra,
exploiting their different shapes in the $k_\parallel - k_\perp$ plane
(see eq. [\ref{approx}]). Exactly how to do so in an optimal fashion when faced with
noisy data deserves further study. In particular, different galaxy types
have a different galaxy bias and number count slope; how to best weigh
their relative contributions to the observed power spectrum?

Lastly, the findings in this paper and paper I cast a new light on
the well known excess correlations seen in pencil beam surveys
\cite{pencil,pencil2}. Could these be the result of enhanced correlations due
to magnification bias, particularly if baryon oscillations are taken
into account (see \cite{eis})? Could the peculiar features seen in the power spectrum
analysis \cite{pencil2} be due in part to the corresponding multiplicative
windows $G$ and $|\tilde W_\parallel |^2$?
We hope to address these questions in the future.

\acknowledgments

We thank Roman Scoccimarro for useful discussions.
LH thanks Ming-Chung Chu and the Institute of Theoretical Physics
at the Chinese University of Hong Kong
for hospitality where part of this work was done.
Research for this work is supported by the DOE, grant DE-FG02-92-ER40699,
and the Initiatives in Science and Engineering Program
at Columbia University. EG acknowledges support from Spanish Ministerio de Ciencia y
Tecnologia (MEC), project AYA2006-06341 with EC-FEDER funding, and
research project 2005SGR00728 from  Generalitat de Catalunya.

\appendix

\section{Incorporating the Alcock-Paczynski Effect}
\label{app:AP}

In this Appendix, we incorporate the Alcock-Paczynski distortion of the correlation function
and power spectrum \cite{AP,matsubara}. The observed redshifts and angles are converted to
comoving radial and transverse distances using the Hubble parameter and angular diameter distance.
Let us denote the parameters used in such a conversion by $\bar H^{\rm AP}$ and ${\bar\chi}^{\rm AP}$. Suppose the true values
for these parameters are ${\bar H}$ and ${\bar\chi}$. Then, the two-point correlation function,
taking into account possible Alcock-Paczynski distortion, is related to the true two-point correlation by
\begin{eqnarray}
\label{xiAP}
\xi_{\rm obs}^{\rm AP} ({\bf x_1}^{\rm AP}; {\bf x_2}^{\rm AP}) = \xi_{\rm obs} ({\bf x_1}; {\bf x_2})
\end{eqnarray}
where the right hand side is the true two-point correlation given in eq. (\ref{obsv}), and
$\delta\chi^{\rm AP}= \delta\chi \bar H/{\bar H}^{\rm AP}$ and
$\delta{\bf x}^{\rm AP}_\perp = \delta{\bf x}_\perp {\bar\chi}^{\rm AP} / \bar\chi$.

Ignoring for the moment redshift distortion due to peculiar motions,
putting eq. (\ref{scaling2}) into eq. (\ref{xiAP}), we have
\begin{eqnarray}
\label{xiAPobs}
&& \xi^{\rm AP}_{\rm obs} (\delta\chi^{\rm AP}, \delta x_\perp^{\rm AP})
= \\ \nonumber
&& \quad \xi_{gg} (\sqrt{ {\delta\chi^{\rm AP}}^2 (1 - \epsilon_H)^2 + {\delta x_\perp^{\rm AP}}^2 (1 - \epsilon_\chi)^2})
\\ \nonumber
&& \quad + (1 - \epsilon_H) \delta\chi^{\rm AP} f ((1-\epsilon_\chi)\delta x_\perp^{\rm AP})
+ g ((1 - \epsilon_\chi)\delta x_\perp^{\rm AP})
\end{eqnarray}
where we have defined
\begin{eqnarray}
\label{epsilonHchi}
1 - \epsilon_H \equiv \bar H^{\rm AP} / \bar H \quad , \quad
1 - \epsilon_\chi \equiv \bar\chi / \bar\chi^{\rm AP}
\end{eqnarray}
The important point is this: the exercise of separating the three
contributions to the observed galaxy correlation,
outlined after eq. (\ref{scaling2})
(see also Fig. 2 in paper I), still works, with
minor modifications. For a fixed $\delta x_\perp^{\rm AP}$, the
magnification bias corrections $(1-\epsilon_H) \delta\chi^{\rm AP} f + g$ still dominate
over $\xi_{gg}$
in the limit of a large $\delta \chi^{\rm AP}$. This allows one to fit for the slope
$(1 - \epsilon_H) f$ and the intercept $g$. 
The cosmology dependent factor of $1 - \epsilon_H$ 
can be absorbed into the galaxy bias factor that is present in $f$, which describes
the galaxy-magnification correlation.
The extrapolation back to $\delta\chi^{\rm AP} = 0$ gives the intercept $g$
which describes the magnification-magnification correlation.
Both $f$ and $g$ are determined up to an overall rescaling of their argument $\delta x_\perp$, 
which is an uncertainty that is always present to the extent
the cosmology dependent angular diameter distance is uncertain.
$\xi_{gg}$ can be obtained by subtracting the inferred
galaxy-magnification and magnification-magnification
contributions from $\xi^{\rm AP}_{\rm obs}$. 

It is common to consider the monopole (eq. 16 in paper I).
Assuming $\epsilon_H, \epsilon_\chi \ll 1$, it can be shown that
\begin{eqnarray}
&& {\rm monopole \,\, of \,\,} \xi^{\rm AP}_{\rm obs} (\delta x^{\rm AP}) =
\\ \nonumber
&& \quad \xi_{gg} (\delta x^{\rm AP} (1 - [\epsilon_H + 2\epsilon_\chi]/3))
\\ \nonumber
&& \quad + (1-\epsilon_H) \delta x^{\rm AP} \tilde f(\delta x^{\rm AP} (1-\epsilon_\chi))
+ \tilde g (\delta x^{\rm AP} (1 - \epsilon_\chi))
\end{eqnarray}
where
\begin{eqnarray}
\tilde f(\delta x^{\rm AP}) \equiv \int_0^{\pi/2} f(\delta x^{\rm AP}{\,\rm sin}\theta_x)
{\,\rm cos}\theta_x
{\,\rm sin}\theta_x d\theta_x \\ \nonumber
\tilde g(\delta x^{\rm AP}) \equiv \int_0^{\pi/2} g(\delta x^{\rm AP}{\,\rm sin}\theta_x)
{\,\rm sin}\theta_x d\theta_x
\end{eqnarray}
and $\theta_x$ is the angle between the separation vector and the LOS.

The appearance of the factor $1 - [\epsilon_H + 2\epsilon_\chi]/3$
in the argument of $\xi_{gg}$ is the origin of the 
common statement that the baryon oscillation scale
measures the combination $({\bar \chi}^2/\bar H)^{1/3}$ \cite{eisenstein}, if one
examines the monopole. Note, however, the presence of the anisotropic
corrections introduced by magnification bias implies the observed monopole
is no longer related to the true monopole by this overall rescaling of $\delta x$.

Let us give the expression for the observed
correlation function in the presence of both peculiar motions and 
the Alcock-Paczynski effect:
\begin{eqnarray}
&& \xi^{\rm AP}_{\rm obs} (\delta {\bf x}^{\rm AP})
= \\ \nonumber
&& \quad \int {d^3 k \over (2\pi)^3} P_{gg}(\bar z, k)
\left(1 + {f_D \over b}{k_\parallel^2 \over k^2} \right)^2 
e^{-{k_\parallel^2 \sigma_z^2}} e^{i {\bf k} \cdot \delta {\bf x}} \\ \nonumber
&& \quad + \, \, qb(1+ \bar z) \int {d^2 k_\perp \over (2\pi)^2}
P_{gm} (\bar z, k_\perp) e^{i {\bf k_\perp} \cdot \delta {\bf x_\perp}} \\ \nonumber
&& \quad \int {d\chi'_1 d\chi'_2\over 2\pi \sigma_z^2} 
e^{-{(\chi_1 - \chi'_1)^2 \over 2\sigma_z^2}} e^{-{(\chi_2 - \chi'_2)^2 \over 2\sigma_z^2}} 
|\chi'_1 - \chi'_2| 
\\ \nonumber 
&& \quad + \, \, (qb)^2 
\int_0^{\bar\chi} d\chi' \left[{(\bar\chi - \chi')\chi' \over \bar\chi}\right]^2 (1+z')^2 \\ \nonumber
&& \quad \int {d^2 k_\perp \over (2\pi)^2} P_{mm} (z', k_\perp) e^{i {\bf k_\perp} \cdot \delta{\bf x}_\perp}
\end{eqnarray}
where $f_D$ and $q$ are as defined in eq. (\ref{fDdef}) and (\ref{qdef}), and
$\delta{\bf x}^{\rm AP}$ and $\delta{\bf x}$ are related as in eq. (\ref{xiAP}).
The first term on the right accounts for both the Kaiser effect and, in a crude form,
the finger-of-god effect due to either virialized motions or photometric redshifts.
The second term on the right represents the contribution from galaxy-magnification
correlation. The finger-of-god dispersion has an effect here, but its effect is
negligible for large LOS separations ($|\chi_1 - \chi_2| \gg \sigma_z$) where the
galaxy-magnification correlation has the largest effect (i.e. 
the integral over $\chi'_1$ and $\chi'_2$ yields roughly $|\chi_1 - \chi_2|
= |\delta \chi|$).
The third term on the right is exactly $\xi_{\mu\mu}$ in eq. (\ref{mumu}): 
the finger-of-god dispersion has no effect on this term, because it has no
intrinsic dependence on the LOS separation $\delta\chi$.
In other words, to good approximation (unless $\sigma_z$ is very large), the
galaxy-magnification and magnification-magnification terms take 
the same form as in eq. (\ref{xiAPobs}). This means the exercise of
separating the three different contributions to the observed correlation
can be repeated here as well.

Lastly, let us give the corresponding
expression in Fourier space.
Defining ${\bf k}^{\rm AP}$, which is related to ${\bf k}$ by
\begin{eqnarray}
\label{kscaling}
k_\parallel^{\rm AP} = k_\parallel \bar H^{\rm AP}/\bar H
\quad , \quad {\bf k_\perp}^{\rm AP} = {\bf k_\perp} \bar\chi /\bar\chi^{\rm AP} \quad ,
\end{eqnarray}
the Alcock-Paczynski distorted power spectrum $P^{\rm AP}_{\rm obs}$ is 
related to the true (windowed) power spectrum $P_{\rm obs}$ by
\begin{eqnarray}
P^{\rm AP}_{\rm obs} ({\bf k}^{\rm AP})
= {\bar H \over \bar H^{\rm AP} } 
{ ({\bar\chi}^{\rm AP})^2 \over {\bar\chi}^2 } 
P_{\rm obs} ({\bf k})
\end{eqnarray}
where $P_{\rm obs}$ is given by
\begin{eqnarray}
&& P_{\rm obs} ({\bf k}) = \\ \nonumber
&& \quad \int {d^3 k' \over (2\pi)^3} P_{gg}(\bar z, k')
\left(1 + {f_D \over b}{(k'_\parallel)^2 \over {k'}^2} \right)^2 \\ \nonumber
&& \quad \quad e^{-{{k'_\parallel}^2 \sigma_z^2}} |\tilde W ({\bf k} - {\bf k'})|^2 \\ \nonumber
&& \quad + \, \, qb(1+ \bar z) G_z (k_\parallel) \int {d^2 k'_\perp \over (2\pi)^2}
P_{gm} (\bar z, k'_\perp) \\ \nonumber
&& \quad \quad |\tilde W_\perp ({\bf k_\perp} - {\bf k'_\perp}) |^2  \\ \nonumber
&& \quad + \, \, (qb)^2 |\tilde W_\parallel (k_\parallel) |^2
\int_0^{\bar\chi} d\chi' \left[{(\bar\chi - \chi')\chi' \over \bar\chi}\right]^2 (1+z')^2 \\ \nonumber
&& \quad \quad \int {d^2 k'_\perp \over (2\pi)^2} P_{mm} (z', k'_\perp) |\tilde W_\perp ({\bf k_\perp} - {\bf k'_\perp} {\chi' /\bar\chi}) |^2
\end{eqnarray}

It is instructive to integrate out the convolving windows following eq. (\ref{Pwinapprox}), and
obtain:
\begin{eqnarray}
\label{PAPaprox}
&& P^{\rm AP}_{\rm obs} ({\bf k}^{\rm AP})
= (1+\epsilon_H) (1+\epsilon_\chi)^2 
P_{gg} (\bar z, k)  \\ \nonumber
&& \Bigl[ \left(1 + {f_D\over b}{k_\parallel^2 \over k^2}\right)^2 e^{-k_\parallel^2 \sigma_z^2} + \\ \nonumber
&& q (1+\bar z) G(k_\parallel) {P_{mm}(\bar z, k_\perp) \over 
  P_{mm}(\bar z, k)} + \\ \nonumber
&& q^2 |\tilde W_\parallel (k_\parallel) |^2 
   \int_0^{\bar\chi} d\chi' (\bar\chi - \chi')^2 (1+z')^2 {P_{mm}(z', k_\perp\bar\chi/\chi') \over
   P_{mm} (\bar z, k)} \Bigr]
\end{eqnarray}
where we have adopted the approximation $G_z \sim G$ as is done in eq. (\ref{approx}).
Note how the distinctive multiplicative windows $G$ and $|\tilde W_\parallel (k_\parallel)|^2$,
which are signatures of the galaxy-magnification and magnification-magnification correlations,
remain intact in the presence of the Alcock-Paczynski effect. 
They give rise to an enhancement in the observed
correlation that is localized to low $k_\parallel$'s.
Just as in the case of the real space correlation function,
quantities such as $P_{gm}$ and $P_{mm}$ (which are related to Fourier
transforms of $f$ and $g$ in eq. [\ref{xiAPobs}]) can be determined
from data only up to a rescaling of their argument
$k_\perp$ (eq. [\ref{kscaling}]) (and in the case of $P_{gm}$, 
an overall normalization which can be absorbed into the galaxy bias).

\newcommand\spr[3]{{\it Physics Reports} {\bf #1}, #2 (#3)}
\newcommand\sapj[3]{ {\it Astrophys. J.} {\bf #1}, #2 (#3) }
\newcommand\sapjs[3]{ {\it Astrophys. J. Suppl.} {\bf #1}, #2 (#3) }
\newcommand\sprd[3]{ {\it Phys. Rev. D} {\bf #1}, #2 (#3) }
\newcommand\sprl[3]{ {\it Phys. Rev. Letters} {\bf #1}, #2 (#3) }
\newcommand\np[3]{ {\it Nucl.~Phys. B} {\bf #1}, #2 (#3) }
\newcommand\smnras[3]{{\it Monthly Notices of Royal
        Astronomical Society} {\bf #1}, #2 (#3)}
\newcommand\splb[3]{{\it Physics Letters} {\bf B#1}, #2 (#3)}

\newcommand\AaA{Astron. \& Astrophys.~}
\newcommand\apjs{Astrophys. J. Suppl.}
\newcommand\aj{Astron. J.}
\newcommand\mnras{Mon. Not. R. Astron. Soc.~}
\newcommand\apjl{Astrophys. J. Lett.~}
\newcommand\etal{{~et al.~}}


\begin{thebibliography}{99}
\bibitem{3Dpaper1} L. Hui, E. Gaztanaga, M. LoVerde, submitted to \prd (2007) [arXiv:0706.1071] (paper I).
\bibitem{matsubara00} T. Matsubara, \apjl 537, 77 (2000).
\bibitem{TOG84} E. L. Turner, J. P. Ostriker, J. R. Gott, \apj 284, 1 (1984);
R. L. Webster,P. C. Hewett, M. E. Harding, G. A. Wegner, Nature  336, 358 (1988);
W. Fugmann, \AaA 204, 73 (1988);
R. Narayan, \apjl 339, 53 (1989); P. Schneider, \AaA 221, 221 (1989).
\bibitem{J95} J. Villumsen, preprint (1995) [astro-ph/9512001].
\bibitem{BTP95} T. J. Broadhurst, A. N. Taylor, J. A. Peacock, \apj 438, 49 (1996).
\bibitem{VFC97} J. Villumsen, W. Freudling, L. N. da Costa, \apj 481, 578 (1997).
\bibitem{MJV98} R. Moessner, B. Jain, J. Villumsen, \mnras 294, 291 (1998)
\bibitem{MJ98} R. Moessner, B. Jain, \mnras 294, L18 (1998).
\bibitem{EGmag03} E. Gazta\~{n}aga,  \apj 589, 82 (2003).
\bibitem{ScrantonSDSS05} R. Scranton, \etal, \apj 633, 589 (2005).
\bibitem{menard} B. Menard, . Bartelmann, \AaA 386, 784 (2002).
\bibitem{bhuv} B. Jain, \apjl 580, 3 (2002).
\bibitem{JSS03} B. Jain, R. Scranton, R. K. Sheth, \mnras 345, 62 (2003).
\bibitem{eisenstein} D. J. Eisenstein \etal, \apj 633, 560 (2005)
\bibitem{2dFa} S. Cole \etal, \mnras 362, 505 (2005)
\bibitem{2dFb} W. J. Percival \etal, preprint (2006) [astro-ph/0608635]
\bibitem{hutsi} G. H\"utsi, \AaA 449, 891 (2006)
\bibitem{tegmark} M. Tegmark \etal, preprint (2006) [astro-ph/0608632]
\bibitem{baotheory} An incomplete list of theoretical papers on the use
of baryon oscillations to measure dark energy includes:
C. Blake, K. Glazebrook, \apj 594, 665 (2003);
E. Linder, \prd 68, 082504 (2003);
W. Hu, Z. Haiman, \prd 68, 063004 (2003);
J. J. Seo, D. J. Eisenstein, \apj 598, 720 (2003);
A. Cooray, \mnras 348, 250 (2004);
T. Matsubara, \apj 615, 573 (2004);
L. Amendola, C. Quercellini, \& E. Giallongo, \mnras 357, 429 (2005);
C. Blake \& S. Bridle, \mnras 363, 1329 (2005).
\bibitem{baoexp} An incomplete list of planned/proposed surveys to measure
the baryon acoustic oscillations includes: 
ADEPT (http://www.jhu.edu/news\_info/news/home06/aug06/ adept.html),
DES Dark Energy Survey (Annis \etal, from {http://decam.fnal.gov}, {http://www.darkenergysurvey.org}),
FMOS/WFMOS at Gemini/Subaru [astro-ph/0507457],
HETDEX ({http://www.as.utexas.edu/hetdex}),
WiggleZ ({http://astronomy.swin.edu.au/~karl/Karl-Home/Home.html}).
\bibitem{scott} A. Vallinotto, S. Dodelson, C. Schmid, J.-P. Uzan (2007) [astro-ph/0702606].
\bibitem{wLHG} M. LoVerde, L. Hui, E. Gazta\~{n}aga, submitted to \prd (2007) [arXiv:0708.0031].
\bibitem{steinmetz} C. Wagner, V. Muller, M. Steinmetz, preprint (2007) [arXiv:0705.0354]
\bibitem{zhangchen} P. Zhang, X. Chen, preprint (2007) [arXiv:0710.1486].
\bibitem{LHGisw} M. LoVerde, L. Hui, E. Gazta\~{n}aga, \prd 75, 043519 (2007) [astro-ph/0611539]
\bibitem{EH98} D. J. Eisenstein, W. Hu, \apj  496, 605 (1998)
\bibitem{smith} R. E. Smith \etal, \mnras 341, 1311 (2003)
\bibitem{asantha} A. Cooray, W. Hu, D. Huterer, M. Joffre, \apjl 557, 7 (2001)
\bibitem{kaiser} N. Kaiser, \apj 498, 26 (1998)
\bibitem{roman} R. Scoccimarro, \prd 70, 083007 (2004)
\bibitem{Hamilton} A. Hamilton, Proceedings of the Ringberg Workshop on Large Scale Structure,
Edited by D. Hamilton, Kluwer Academic, Dordrecht (1996) [astro-ph/9708102]
\bibitem{AP} C. Alcock, B. Paczynski, Nature 281, 358 (1979)
\bibitem{matsubara} 
T. Matsubara, Y. Suto, \apjl 470, 1 (1996);
T. Matsubara, A. S. Szalay, \prl 90, 21302 (2003)
\bibitem{sss} R. E. Smith, R. Scoccimarro, R. K. Sheth, preprint (2006) [astro-ph/0609547]
and (2007) [astro-ph/0703620]
\bibitem{pencil} T. J. Broadhurst, R. S. Ellis, D. C. Koo, A. S. Szalay, Nature 343, 726 (1990).
\bibitem{pencil2} A. Szalay, T. J. Broadhurst, N. Ellman, D. C. Koo, R. S. Ellis,
Proc. Natl. Acad. Sci. 90, 4853 (1993).
\bibitem{eis} D. J. Eisenstein, W. Hu, J.Silk \& A.Szalay  \apjl  494, L1 (1998)

\end{thebibliography}
\end{document}